\documentclass[fleqn,10pt]{wlscirep}

\usepackage[utf8]{inputenc}
\usepackage[T1]{fontenc}

\usepackage{enumitem}
\usepackage{color,soul}
\usepackage{subcaption}
\title{The Skincare project, an interactive deep learning system for differential diagnosis of malignant skin lesions. Technical Report}

\author[1]{Daniel Sonntag}
\author[1]{Fabrizio Nunnari}
\author[1]{Hans-J\"urgen Profitlich}
\affil[1]{German Research Center for Artificial Intelligence, Saarbr\"ucken, Germany}

\begin{abstract}
A shortage of dermatologists causes long wait times for patients who seek dermatologic care from general practitioners. In addition, the diagnostic accuracy of general practitioners has been reported to be lower than the accuracy of artificial intelligence software. 
This article describes the Skincare project (H2020, EIT Digital). Contributions include enabling technology for clinical decision support based on interactive machine learning, a reference architecture towards a Digital European Healthcare Infrastructure (also cf. EIT MCPS), technical components for aggregating digitised patient information, and the integration of decision support technology into clinical test-bed environments. However, the main contribution is a diagnostic and decision support system in dermatology for (1) patients and (2) doctors, an interactive deep learning system for differential diagnosis of malignant skin lesions. In this article, we describe its functionalities and the user interfaces to facilitate machine learning from human input. The baseline deep learning system, which delivers state-of-the-art results and the potential to augment general practitioners and even dermatologists,  was developed and validated using de-identified cases from a dermatology image data base (ISIC), which has about 20000 cases for development and validation, provided by board-certified dermatologists defining the reference standard for every case. ISIC allows for differential diagnosis, a ranked list of eight diagnoses, that is used to plan treatments in the common setting of diagnostic ambiguity. We give an overall description of the outcome of the Skincare project (2019 and 2020), and we focus on  the steps to support communication and coordination between humans and machine in interactive machine learning. This is an integral part of the development of future cognitive assistants in the medical domain, and we describe the necessary intelligent user interfaces. Future work will be needed to prospectively assess the clinical impact of using this interactive tool extension in actual clinical workflows.
\end{abstract}

\begin{document}

\flushbottom
\maketitle

\thispagestyle{empty}

\section{Introduction}

Recent advances in deep learning have facilitated the development of artificial intelligence tools to assist in diagnosing skin cancer from dermatoscopic images. We combine patient records with images from the smartphone without a dermatoscope (macroscopic images)  and  dermatoscopic images with a dermatoscope (microscopic images)  for knowledge discovery and knowledge acquisition toward decision support and services in clinical and non-clinical environments. Input modes include smartphones for a direct digitisation of patient data and images. Prior works have focused on the visual recognition of skin lesions from dermoscopic images, which requires a dermatoscope. The innovative aspect is a holistic view on individual patients based on teledermatology, whereby patient data and lesions photographed with a mobile device can be taken into account for clinical decision support. In order to provide such a system, many technical questions around machine learning and the interaction with the domain user have to be answered. The starting point is a system based on microscopic (dermoscopic) images, to be augmented by macroscopic images after the deployment of the system.  Our baseline visual classification system provides a differential diagnosis across 8 conditions: 
"Melanoma",
"Melanocytic nevus",
"Basal cell carcinoma",
"Actinic keratosis",
"Benign keratosis (solar lentigo / seborrheic keratosis / lichen planus-like keratosis)",
"Dermatofibroma",
"Vascular lesion", and 
"Squamous cell carcinoma". The images and classifications stem from The International Skin Imaging Collaboration (ISIC); this is a combined academia and industry effort aimed at improving melanoma diagnoses and reducing melanoma mortality by facilitating the application of digital skin imaging technologies, see  \url{https://dermoscopedia.org/ISIC_project}. 
 
Image labelling enabled by explainable deep learning technology should enhance dermatologists' interactions with such instances of interactive machine learning (IML) systems in the future. We give an overall description of DFKI's main outcome of the Skincare project, focussing on its functionalities, and the user interfaces to facilitate machine learning from human input: section~\ref{sec:objectives} lists the objectives of the project in more detail. Section~\ref{sec:functionalities} gives an overview of the functionalities offered by the system. Section~\ref{sec:architecture} illustrates the overall software architecture of the system. Section~\ref{sec:user-interfaces} describes the end-user interface implemented to enable a stand-alone and interactive use of the decision support and interactive machine learning system. Section~\ref{API} describes the REST API of the intelligent user interface's web page.

\section{Objectives}
\label{sec:objectives}

The goal of the project Skincare (see \url{http://ai-in-medicine.dfki.de/}) is to develop a decision support system for the diagnosis of malignant skin lesions (e.g., Melanoma). Contributions include enabling technology for clinical decision support based on interactive machine learning, a reference architecture towards a Digital European Healthcare Infrastructure (also cf. EIT MCPS), technical components for aggregating digitised patient information (patient record and teledermascopic images), and the integration of decision support technology into clinical test-bed environments. However, the main contribution is a diagnostic and decision support systems in dermatology towards an interactive deep learning system for differential diagnosis of malignant skin lesions.

The software should support the dermatologists on several aspects: i) providing a colour analysis of the image, ii) highlighting regions of interest (RoIs) on the images, iii) proposing a differential diagnosis. The interactive learning system runs on top of this decision support system, to elicit user feedback via a graphical user interface (GUI) for improving the machine learning model.  However,  core functionalities of the system must be GUI-agnostic and available for all the partners of the project. They will be made available as a stand-alone web service accessible through a REST API.
On top of that, a web-based GUI will allow dermatologists to interactively use and improve the system.

From a machine learning perspective, we provide an environment for grounding decisions based on visualisations of skin anomalies (and the internal workings of the machine learning systems) to identify and visualise such anomalies. More precisely, and similar to \cite{yan19} we create attention maps to provide  more interpretable output as opposed to only outputting a classification label of the whole image without internal structure. In addition, the segmentation explained in section \ref{sec:functionalities} delivers medical visual prior information (as opposed to background information about the patient) to boost the classification performance and deliver better localisations of attention maps indicating specific regions of the images. This should help, in the end, to gain insights into which image regions contribute to the results. A further question is how to train future models to focus on the expected regions of interest (ROIs) and validate the effectiveness of ROI priors. 

Focussing on the visual analytics and machine learning part, the objective is to support model explanation and interpretation for deep learning decisions based on ROIs relevant for classification. Model debugging and improvement are additional objectives. In addition to the general aim of democratising machine learning for non-experts, we  tackle the challenge of interpretability and transparency of blackbox models for the medical expert. A general sense of model understanding should be essential for medical decision support. A recent survey on visual analytic and deep learning classifies such systems around attention maps and heatmaps for image classification \cite{hohman18}. Our recent survey at DFKI focusses on actual tools for visual analytics in machine learning in addition. Here, an interpretation is a special form of explanation, where we show the predictions by elucidating the (spatial) mechanism by which the model works. This interpretation is the mapping of an image and its classification into a domain (e.g., the locations-based heatmap) the medical expert can make sense of. This interpretation of the concepts interpretability and explainability are in line with recent definition in the scientific community \cite{lipton18, MONTAVON18}. Also, insights for social sciences suggest that explanations in AI-based decision making processes should have a structure that the domain experts can accept \cite{MILLER19};  location-based heatmaps as used here are not controversial to how doctors explain  malignant skin
lesions to each other and can serve as a useful starting point for explaining computer-aided differential diagnosis in the medical domain.

In the Skincare project, we visualise feature spaces by using visual detectors of ROIs. By visualising ROIs of correctly classified image instances (or misclassified ones), we gain insight and an intuitive understanding of how the dermatology recognition system works.  
The identification of the ROIs is performed using algorithms fostering the explainability of decisions taken by deep learning approaches. In image classification problems solved with CNNs, identifying ROIs corresponds to identifying the groups of pixel that contribute to the final classification choice. The granularity of the groups depends on the explainability technique. Each group is also associated to a quantification of the contribution, called \emph{saliency}. The output of an explanation algorithm is called \emph{saliency map}, which is often converted into a \emph{heat map}, via a direct gray-scale-to-color scheme conversion, for both aesthetic reasons and an easier human interpretation.

The algorithms for attributing important regions and feature visualisation either highlight important regions of the image (attribution) or create a heatmap that is representative of the features for a given class, e.g., melanoma. We generate translucent heatmaps in real-time as overlays to highlight important regions that contribute to the AI-based classification. It is important  to note that we visualise the deep learning process after training, hence we perform a dermatology image instance-based analysis and exploration; a further interactive experimentation task is explained further down. For the Skincare project, we use two visual explanation techniques with translucent overlays: GradCAM \cite{selvaraju_grad-cam:_2017} and RISE \cite{petsiuk_rise:_2018}.
GradCAM computes heatmaps through an analysis of the CNN structure, namely, by measuring the intensity of the inner activation values at a given convolution stage. Complementarily, RISE uses a stochastic method; an input image is repeatedly classified while being bombed with masking noise (grid of black quadrants) at random frequencies and phases, while the contribution of each pixel to the classification of the input sample is accumulated.

The GradCAM algorithm is very fast and memory-saving, as it requires only a single pass of forward and back-propagation through the network, but the final resolution of the saliency map is limited by the resolution of the last convolutional layer.
RISE can potentially reach very high image resolutions, but its applicability is limited by the quantity of computer memory available for the accumulation matrix and the time needed to repeat multiple (likely more than 1000) predictions to achieve reliable results. Both methods can be triggered via the web page as shown in figures \ref{fig:explanation_rise} and \ref{fig:explanation_gradcam}.

\section{Functionalities}
\label{sec:functionalities}

The Skincare system allows the user to interact with a classifier to detect malign lesions. After uploading dermatological images, the user can start an analysis which delivers the following results (see also figures \ref{fig:classification} and \ref{fig:classificationtouch}):

\begin{itemize}
    \item a binary classification of images between benign or malign or an 8-class classification (partly based on the following segmentation and feature extraction steps);   
    \item a segmentation of the lesion, i.e., a separation of the areas pertaining to the lesion from the rest of the healthy skin. One version of the segmentation is implemented by a neural network (pixel-oriented); a second version is implemented by visual image preprocessing techniques (visual filter-oriented colour analysis). 
    \item feature extraction: the identification of five types of dermoscopic feature patterns: globules, streaks, pigment network, milia-like cyst, and negative network;
    \item feature extraction: numerical indicators for borders, diameters, and asymmetries;
    \item feature extraction: graphical features supporting the ABCD rule including segmentation, color areas, and asymmetries.
\end{itemize}

Additionally, the user can upload sets of benign and malign images and evaluate the classification results under different thresholds for malignity. A wide number of quality measures are computed and presented graphically.

\subsection{Binary Classification (Malignant vs. Benign)}

\begin{figure}
	\centering
	\includegraphics[width=1.0\textwidth]{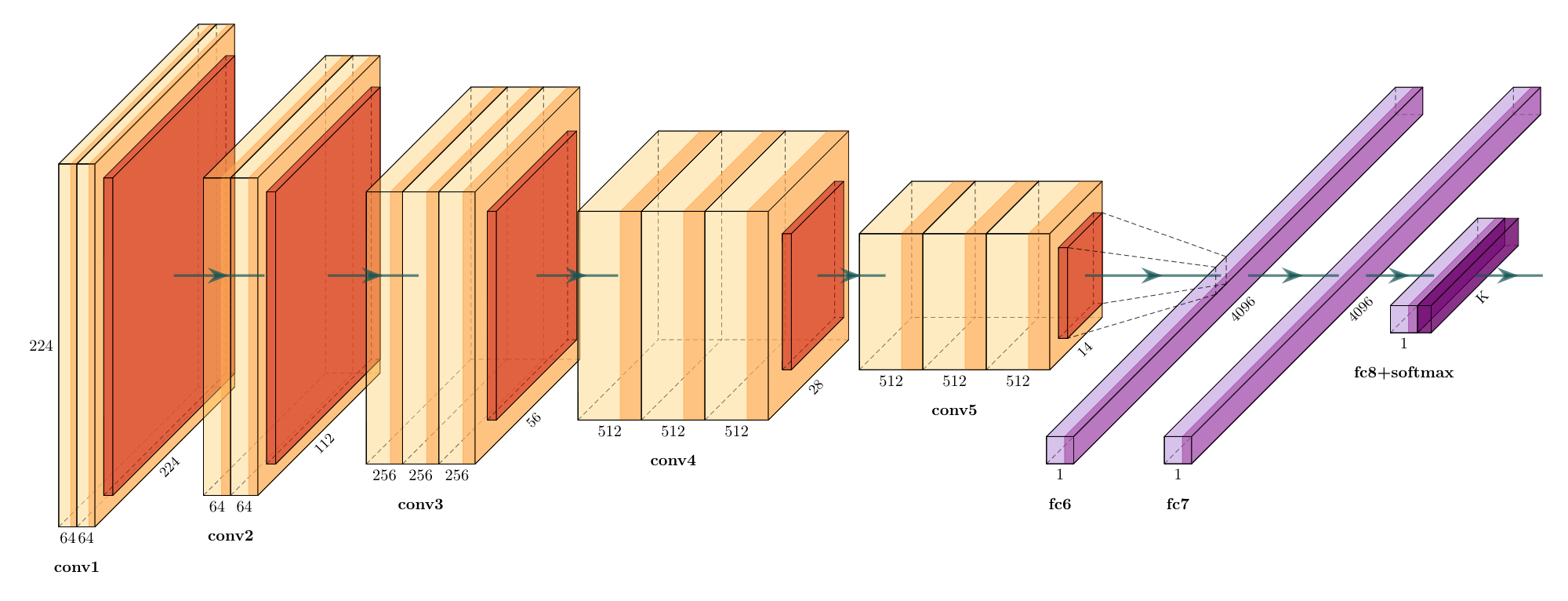}
	\caption{Basic architecture of the VGG16 CNN for image classification}
    \label{fig:vgg16-arch}
\end{figure}

Classification is the task of assigning an input image to one of several pre-defined categories. 
The classifier is based on a VGG16\footnote{\url{http://www.robots.ox.ac.uk/~vgg/research/very_deep/}} \cite{simonyan_very_2014} deep convolutional neural network (see figure \ref{fig:vgg16-arch}) trained on a set of about 25000 lesion images downloaded from the open ISIC repository in February 2019.
The architecture has 5 convolution blocks, each appended with 2x2 max pooling layer. The convolution blocks are followed by 2 fully connected layers (fc1, fc2) followed by a softmax (sigmoid+multinomial) activation instead of the original ReLU activation. We tried different number of neurons in fc1 and fc2 layers, ranging from 512 to 4096 in powers of 2, and found that 2048 gives the best performances.

The network has been trained by applying transfer learning from the original version, which was trained for the classification among 1000 classes on the ImageNet dataset \cite{deng_imagenet:_2009}.
As common practice in transfer learning, we kept the pre-trained  weights of the convoutional layers and modified the final part of the network to accommodate the classification among a reduced set of classes (currently two, malignant or benign, or the eight classes mentioned in the introduction). As a preprocessing step, images are cropped and scaled to a resolution of 450x450 pixels.
During learning, we used stochastic gradient descent (SGD) with Nesterov acceleration. The learning rate is 0.0001 with 0.9 momentum. We used the weight decay policy with coefficient 0.0001. We performed a 48-times data augmentation by both flipping and rotating each image 24 times (15 degree steps), as suggested by Fujisawa et al. \cite{fujisawa_deep-learning-based_2018}. Additional details about the CNN structure, training parameters, and performances can be found in \cite{nunnari_cnn_2019}. What follows is the description of the 8-class classification. Future extensions will provide a more fine-grained classification of the most common lesion types of the medical literature.

\subsection{8-class Classification}

Our baseline visual classification system provides a differential diagnosis across 8 conditions: 
"Melanoma (MEL)",
"Melanocytic nevus (NV)",
"Basal cell carcinoma (BCC)",
"Actinic keratosis (AK)",
"Benign keratosis (BKL)",
"Dermatofibroma (DF)",
"Vascular lesion (VASC)", and 
"Squamous cell carcinoma (SCC)".
The model is trained using the same dataset provided from the ISIC2019 challenge\footnote{\url{https://challenge2019.isic-archive.com/}}.
The distribution among the classes is reported at the top of table \ref{tab:8cls-count}.

The classification of the 8 types of lesions is performed by the same architecture used for binary classification (see figure \ref{fig:vgg16-arch}), with the difference in the number of outputs.

For these models, we used $\sim$20k images for training, and the remaining images are equally divided for validation and test.
Training the Model-8cls for 10 epochs using an 9th gen 8-core i7 CPU and an NVIDIA RTX TITAN GPU (24 GB RAM) took about 3 days and 17 hours. With the available GPU memory we could train with a batch size of 16.

\begin{table}[t]
    \centering
    \footnotesize
\begin{tabular}{l|rrrrrrrr|r}
\toprule
Class &  MEL & NV & BCC & AK & BKL & DF & VASC & SCC & Tot \\
\midrule
Count & 4346 & 10632 & 3245 & 845 & 2333 & 235 & 222 & 622 & 22480 \\
Pct & 17.8\% & 50.8\% & 13.1\% & 3.4\% & 10.4\% & 1.0\% & 1.0\% & 2.5\% & 100\% \\
\bottomrule
\end{tabular}
    \caption{Class distribution in the ISIC 2019 training set.}
    \label{tab:8cls-count}
\end{table}

\begin{table}[t]
    \centering
    \footnotesize
\begin{tabular}{l|cccccccc}
\toprule
Lesion & MEL & NV & BCC & AK & BKL & DF & VASC & SCC\\
Frequency &17.8\% & 50.8\% & 13.1\% & 3.4\% & 10.4\% & 1.0\% & 1.0\% & 2.5\% \\
\midrule
Accuracy  & 0.844 & 0.824 & 0.925 & 0.962 & 0.918 & 0.985 & 0.995 & 0.970 \\
Specificity & 0.892 & 0.895 & 0.953 & 0.975 & 0.951 & 0.986 & 0.996 & 0.977 \\
Sensitivity & 0.626 & 0.756 & 0.738 & 0.570 & 0.634 & 0.783 & 0.880 & 0.694 \\
ROC AUC & 0.867 & 0.920 & 0.960 & 0.955 & 0.921 & 0.964 & 0.991 & 0.953\\
\bottomrule
\end{tabular}
    \caption{Performance metrics of the 8-class classification model.}
    \label{tab:8cls-test}
\end{table}

Table \ref{tab:8cls-test} shows the state-of-the-art results on the test set ($\sim$1.5k samples) in terms of Accuracy, Specificity, Sensitivity, and ROC AUC, together with the distribution of image samples.
In general we can observe that the model performs better in specificity (up to 0.996 for the VASC class) rather than sensitivity (with a lower value of 0.570 for class AK).
Interestingly, better results are obtained for classes with less samples than for classes with more samples, probably because of the specific morphology of such pathologies.

\subsection{Presentation of classification results (Confidence)}

Given an image to classify, neural networks output a \emph{probability distribution}, i.e., they associate a real number between 0 and 1 to each of the classes for which the network is trained. The numbers, i.e., probabilities ($p$), sum up to $1.0$.
Typically, the results of the classification are presented to a user by selecting the class with the highest $p$-value, or by selecting a fixed-size list of the top $k$ images with the highest $p$-values. 

However, this method of presenting the classification results can lead to erroneous interpretation about the \emph{reliability} or \emph{confidence} of the classification. 
For example, consider a binary classification between benign or malignant lesions. Given an image, the system returns a probability distribution of $p=0.51$ for benign and $p=0.49$ for malignant. It is common practice to simply select the class with the highest $p$-value and inform the users that the analysis resulted in a verdict for benign, thus hiding the ambiguity of the decision.
Of course, presenting the user with all the $p$-values would give him or her more detailed information. However, these are harder to interpret when performing a classification on multiple classes, where $p$-values for non-top-1 classes can be small numbers, very close to 0.

Hence, we use a methodology which presents a variable number of top images according to a level of \emph{confidence} $c$, which is computed on the $p$-value as the percentage on the range between a uniform distribution threshold $u$ (1.0 divided by the number of classes) and 1.0.
Back to the previous example on 2 classes, $p=0.51$ would lead to $c=0.5\%$ confidence, $p=0.75 \rightarrow c=50\%$, and $p=1.0 \rightarrow c=100\%$. More formally, given $n$ classes:
\begin{equation*}
c=\frac{p-u}{1.0-u}*100
\end{equation*}
where $u=1/n$, $n$ is the number of classes, and images are presented to the user only if $p>u$.
The rationale behind this is to discard all images whose $p$ is below the threshold of pure uniform probability distribution (total uncertainty).
For example, consider an 8-class classification problem (where $u=0.125$) where an image gets $p=0.67$ for Benign Keratosis, $p=0.28$ for Melanoma, and $p<0.125$ for all other classes.
The system will show to the user that the classification is likely to be Benign Keratosis with confidence $62.4\%$, but also likely a Melanoma with confidence $17.4\%$.

\subsection{Segmentation (with CNN)}
\label{sec:dlsegmentation}
\begin{figure}
    
    \hfill
    \mbox{\begin{subfigure}[t]{0.25\textwidth}
    \centering\fbox{\includegraphics[width=\textwidth]{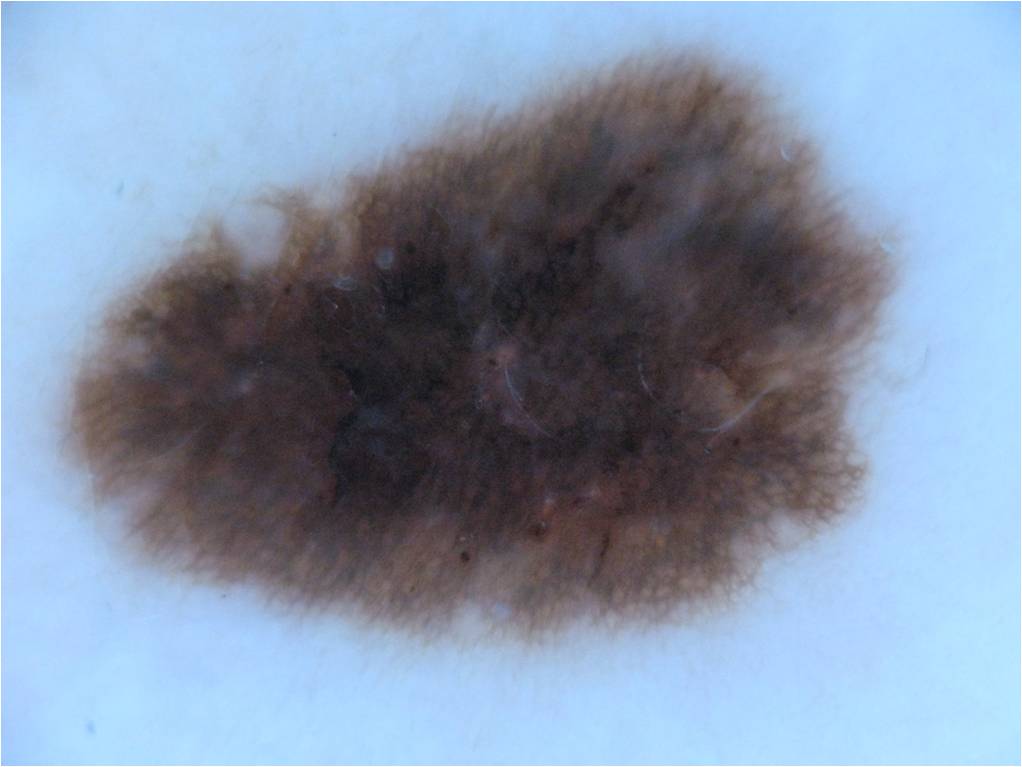}}
    \subcaption{original}
    \end{subfigure}}\hfill
    \mbox{\begin{subfigure}[t]{0.25\textwidth}
    \centering\fbox{\includegraphics[width=\textwidth]{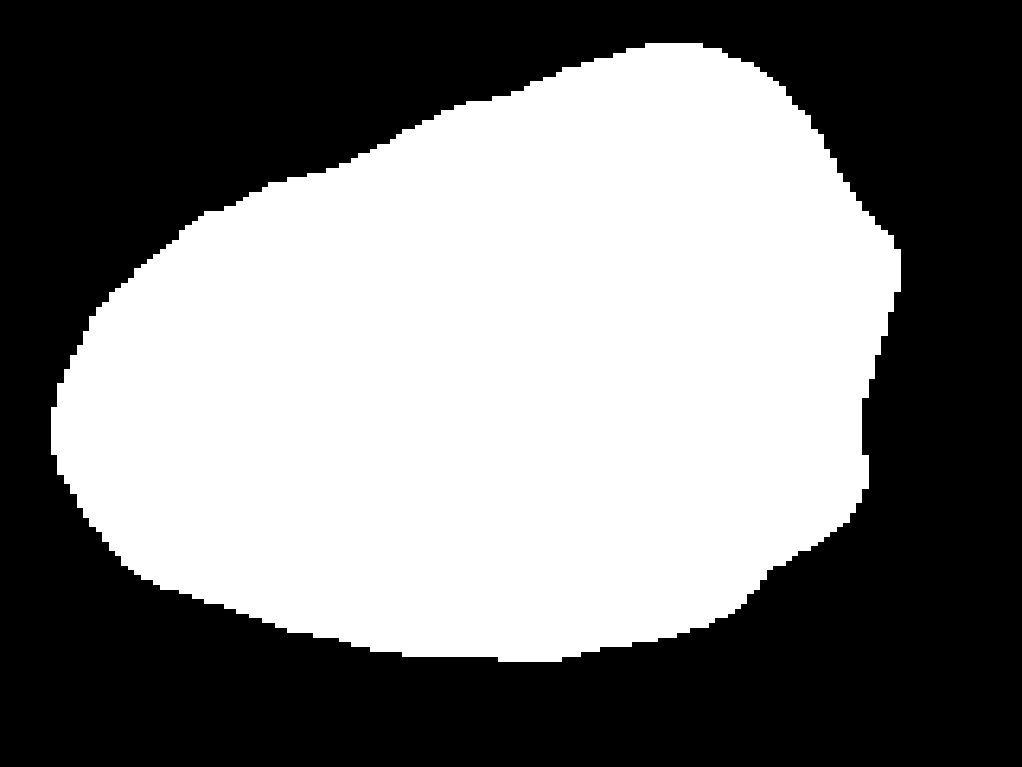}}
    \subcaption{segmentation}
    \end{subfigure}}\hfill
    \mbox{\begin{subfigure}[t]{0.25\textwidth}
    \centering\fbox{\includegraphics[width=\textwidth]{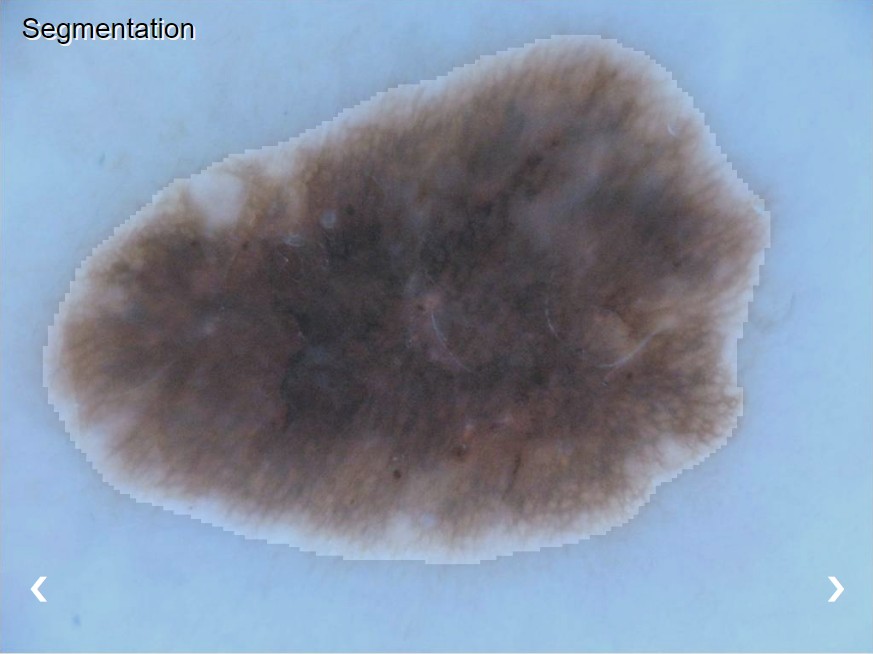}}
    \subcaption{segmentation overlay (0.9)}
    \end{subfigure}}\hfill\hfill
    
    \caption{Segmentation: a) original image, b) DL-based segmentation, c) overlay} 
    \label{fig:dl_segmentation}
\end{figure}

Segmentation is the task of tracing a contour on the image in order to separate the lesion area of the skin from the healthy part. The segmentation task is performed by a masking model based on the U-Net architecture \cite{navab_u-net:_2015}, which is a (de-)convolutional deep neural network for pixel-level classification.
The model was trained using data from the ISIC 2018 Task 1, where the input are 2594  skin lesion images  in RGB colour format (see figure \ref{fig:dl_segmentation}a) and, for each sample, the ground truth is a binary mask of the same resolution of the input image (e.g., figure \ref{fig:dl_segmentation}b). The ground truth masks have been created by filling the manually-traced contour of lesions (by dermatologists). Hence, a white pixel in the mask indicates a pixel pertaining to the lesion in the corresponding RGB image, while black pixels are associated with the  surrounding skin.

\begin{figure*}[t]
	\centering
	\includegraphics[width=1.0\textwidth]{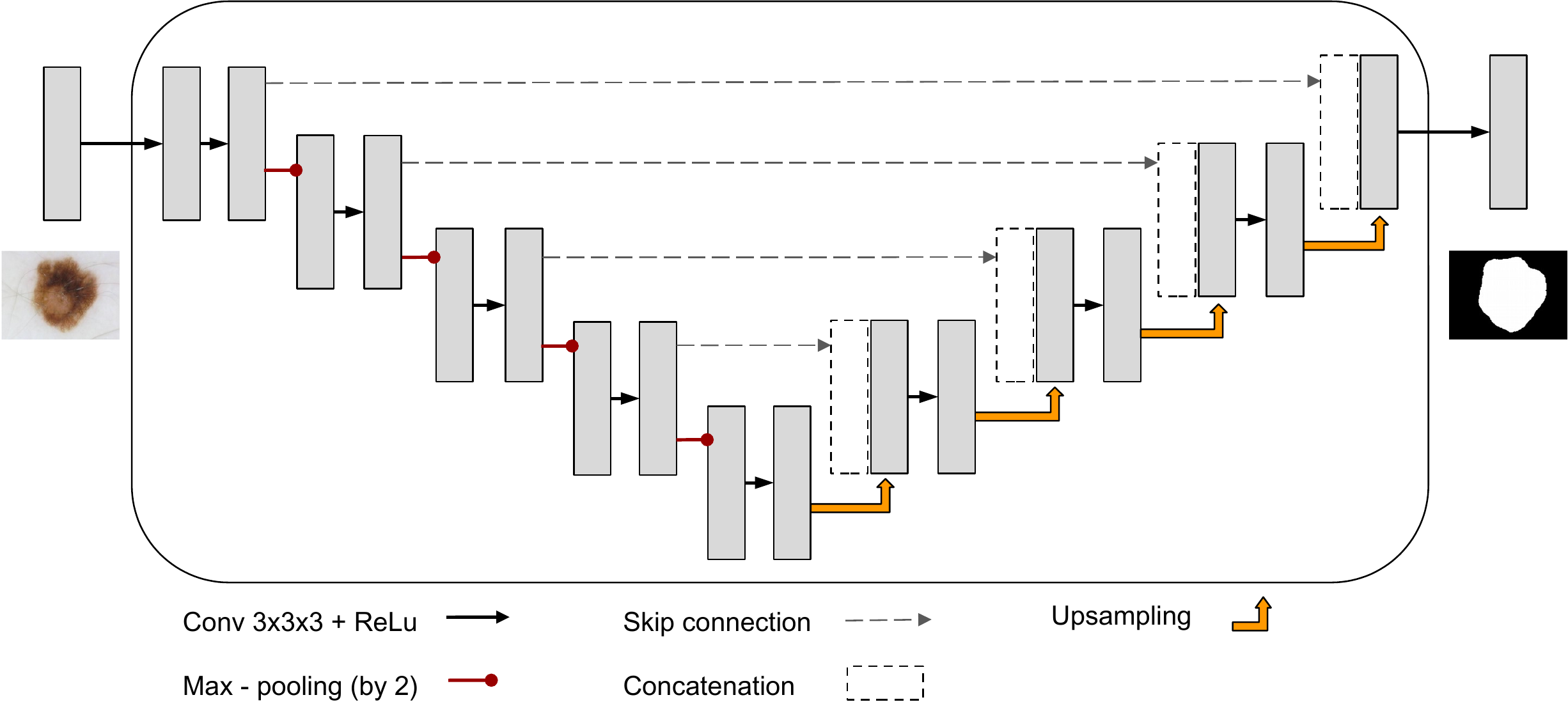}
	\caption{The architecture of the U-Net used for lesion segmentation. The input image is 3-channels RGB, while the output image is 1-channel gray-scale with the same resolution.}
	\label{fig:u-net}
\end{figure*}

Figure \ref{fig:u-net} shows the U-Net architecture together with a sample input and output.
The architecture is composed of 9 convolution blocks, each of them is a pair of 2D \emph{same} convolutions with stride 3x3x3.
Downsampling is the result of a max-pooling with size 2x2.
Upsampling is the result of a 2x2 transposed 2D \emph{same} convolution. After each upsampling step, the convolution is performed on the concatenation between the upsampling result and the output of the downsampling with corresponding resolution.
The initial number of filters (32) doubles at each downsampling.
For this work, we used an input/output resolution of 180x180 pixels.

To train the U-Net we randomly sampled 20\% of the total 2594 images as a test set, and extracted another 5\% of the remaining samples for a final distribution of 2075/104/415 images for training/validation/test.
Images in the training set were pre-processed to center data by subtracting the mean per channel and scaled to match the input resolution of U-Net.
After prediction, the output images are converted into binary masks using a threshold of 0.5.

We trained the U-Net for $100$ epochs, optimising for cross-entropy loss, reaching the lowest loss after 40 epochs, and obtained a Jaccard index J=0.75.
The Jaccard index constrains the generated lesion boundaries to be spatially and geometrically precise invariant of lesion contours; it computes the expectation of pixel-wise similarity between a predicted image segmentation and its corresponding ground truth as a measure of the intersection over union. The Jaccard Index is defined as:
\begin{equation}
J = \frac{\sum y_{truth}\,y_{predict} }{\sum y_{truth}^{2} + \sum y_{predict}^2 - \sum y_{truth}\,y_{predict} }
\end{equation}
where $y_{truth}$ and $y_{predict}$ represent the ground truth and predicted pixel values respectively, with $y \in \{0,1\}$, with sums taken over the dimension of the image.

The current implementation computes segmentation masks of a  resolution of 180x180 pixels, which are then re-scaled to the resolution of the original image for overlapping and interactive masking (see figure \ref{fig:dl_segmentation}c with overlay 0.9).

\subsection{Segmentation (visual filter-oriented colour analysis)}

Dermoscopic images may contain artefacts, such as hairs, ruler markings, air bubbles, and uneven illumination. The first stage in this segmentation pipeline towards classifying  involves some preprocessing of the captured image, in order to remove the effects of such artefacts, reduce noise, and enhance the image contrast in the image. In this preprocessing step, color transformation is performed first, where the RGB image is converted to the Y’UV color space that separates the color (chroma) component from the illumination component of the image. This conversion helps the classification process  to perform consistently under varying illumination conditions. 

To reduce unwanted effects of further artefacts, the image is smoothed by a two-dimensional (2-D) Gaussian filter \cite{9pmid15691255, 19DBLP:conf/iciar/MajtnerYH16, 21DBLP:journals/sj/BarataRFMM14,  22DBLP:journals/sj/BarataRFMM14}, that produces an image by performing convolutions of the filter with the image. The size of the filter is determined by its kernel value $k$, where large kernel values significantly blur the image and weaken the borderline along with noise, whereas small kernel values do not reduce the noise to a desirable extent. Tests showed that a kernel of $k=5$ and a standard deviation of $\sigma=1$ provide the best results. Alternative filters are median \cite{16pmid25571546, 23pmid27265054, 24pmid20832992}, and anisotropic diffusion filters \cite{25DBLP:journals/eswa/OliveiraMPT16,26DBLP:journals/amc/BarcelosP09}.

\begin{figure}[!htb]
    
   \mbox{\begin{subfigure}[t]{0.2\textwidth}
    \centering\fbox{\includegraphics[width=\textwidth]{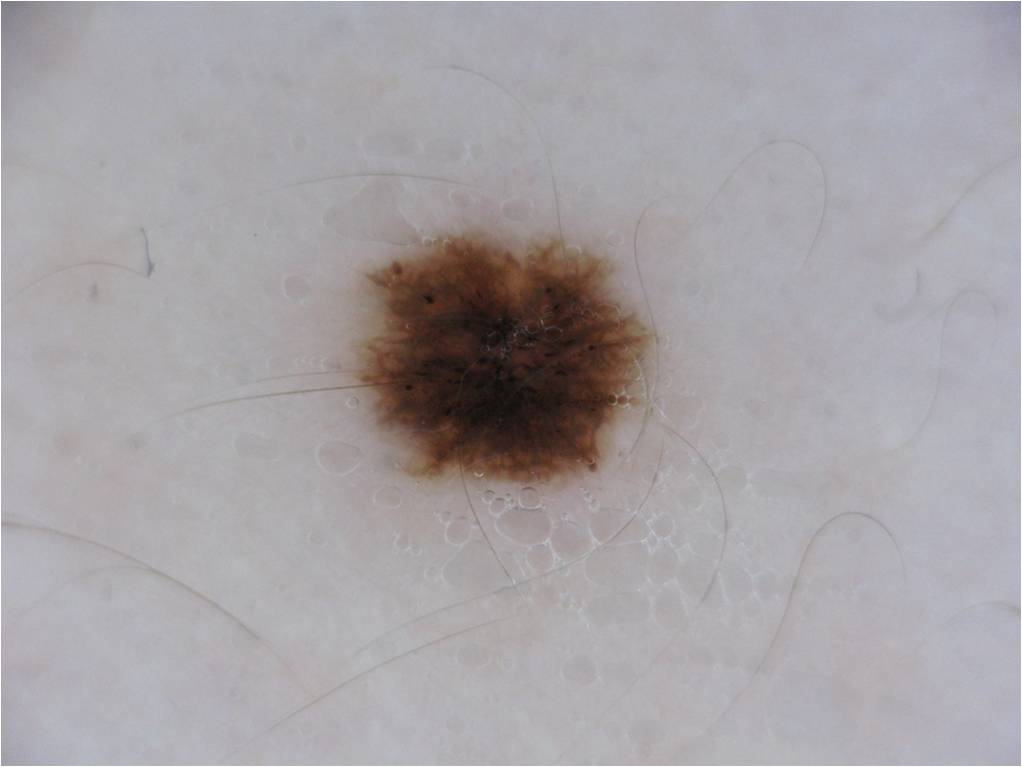}}
    \subcaption{}\end{subfigure}}\hfill
    \mbox{\begin{subfigure}[t]{0.2\textwidth}
    \centering\fbox{\includegraphics[width=\textwidth]{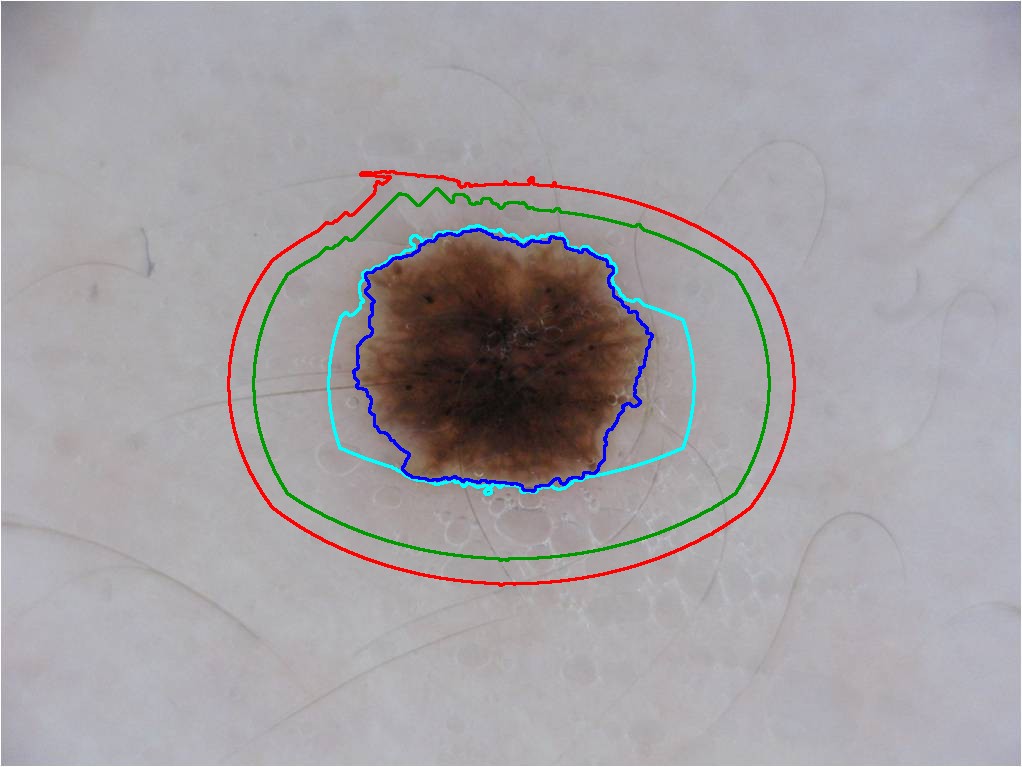}}\subcaption{}
    \end{subfigure}}\hfill
    \mbox{\begin{subfigure}[t]{0.2\textwidth}
    \centering\fbox{\includegraphics[width=\textwidth]{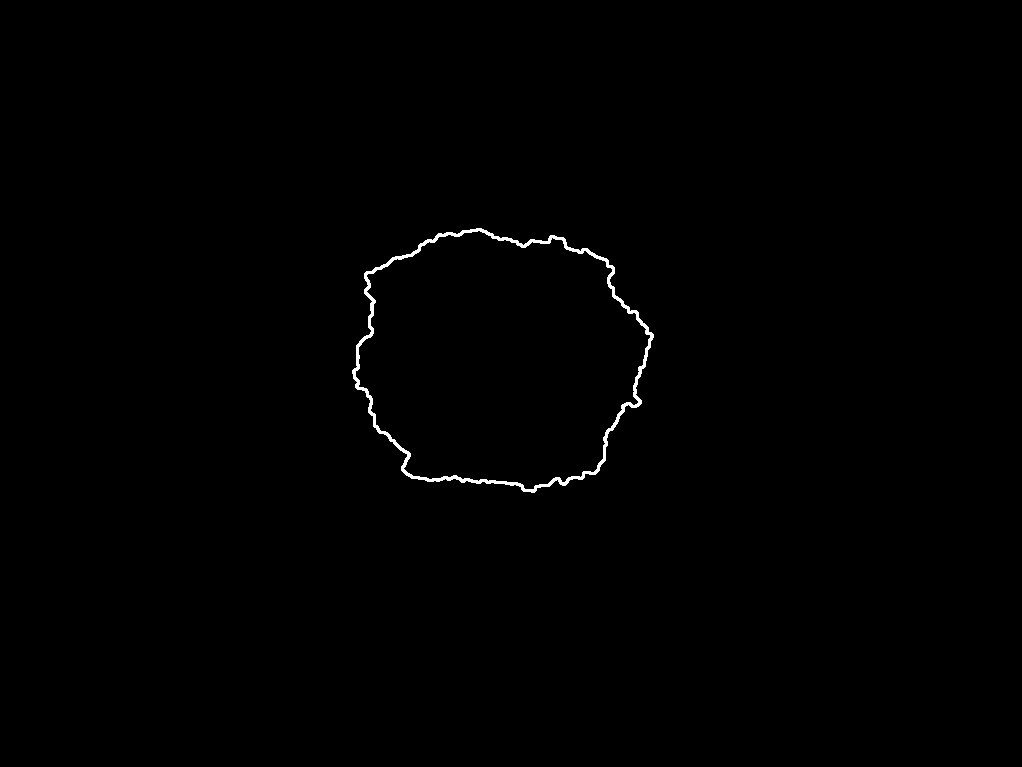}}\subcaption{}
    \end{subfigure}}\hfill
    \mbox{\begin{subfigure}[t]{0.2\textwidth}
    \centering\fbox{\includegraphics[width=\textwidth]{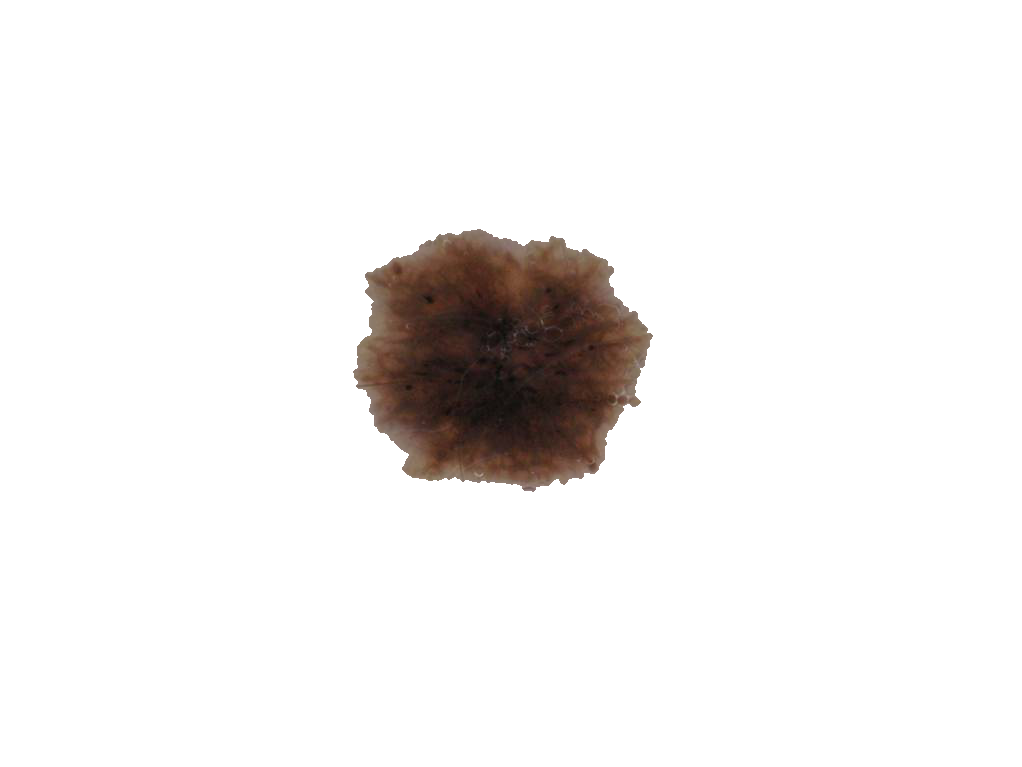}}
    \subcaption{}
    \end{subfigure}}
    
    \caption{Segmentation steps: a) original image, b) active contours evolution steps, c) active contours result, d) resulting segmentation of the lesion.}\label{fig:steps_segmentation}
\end{figure}

A number of segmentation algorithms have been reported in the literature, such as the edge detection \cite{26DBLP:journals/amc/BarcelosP09, 28pmid27215953}, thresholding \cite{30pmid25585429}, and active contour methods. Some segmentation algorithms are very sensitive to noise and require high-contrast images, sometimes in addition to input from the user to adjust the segmented region. Recently active contour algorithms have come up, where a flexible curve narrows until it fits the boundary of the ROI. 
Active contour algorithms can be categorised as parametric or geometric based on the curve tracking method. Parametric models have difficulties in handling ROIs with large curvatures and topological changes \cite{10pmid18003531,23pmid27265054}. The geometric active contour model improves by adapting to topological changes. Generally speaking, active contour models involve solving partial differential equations (PDEs) for curve evolution, creating a computational burden \cite{33pmid18390371}. One popular geometric active contour model is known as the Chan–Vese model \cite{25DBLP:journals/eswa/OliveiraMPT16, 32pmid18249617}. SkinCare uses a modified Chan–Vese model that runs in real-time with fast level-set approximation \cite{34SutaBessyVeja2012}. The model updates the curve during each evolution until it fits the boundary of the object of interest. Figure \ref{fig:steps_segmentation} shows the processing steps from the original image (a), shrinking surrounding circles (b), the resulting contour of the ROI, to the final segmented lesion.
This computation of segmentation is completely based on computer vision algorithms whereas the segmentation described in section \ref{sec:dlsegmentation} is based on a pixel-wise CNN mask.

\subsection{Feature Extraction (U-Net)}
\label{fe}

\begin{figure}
    \hfill
    \mbox{\begin{subfigure}[t]{0.2\textwidth}
    \centering\fbox{\includegraphics[width=\textwidth]{images/original.jpg}}
    \subcaption{original}
    \end{subfigure}}\hfill
    \mbox{\begin{subfigure}[t]{0.2\textwidth}
    \centering\fbox{\includegraphics[width=\textwidth]{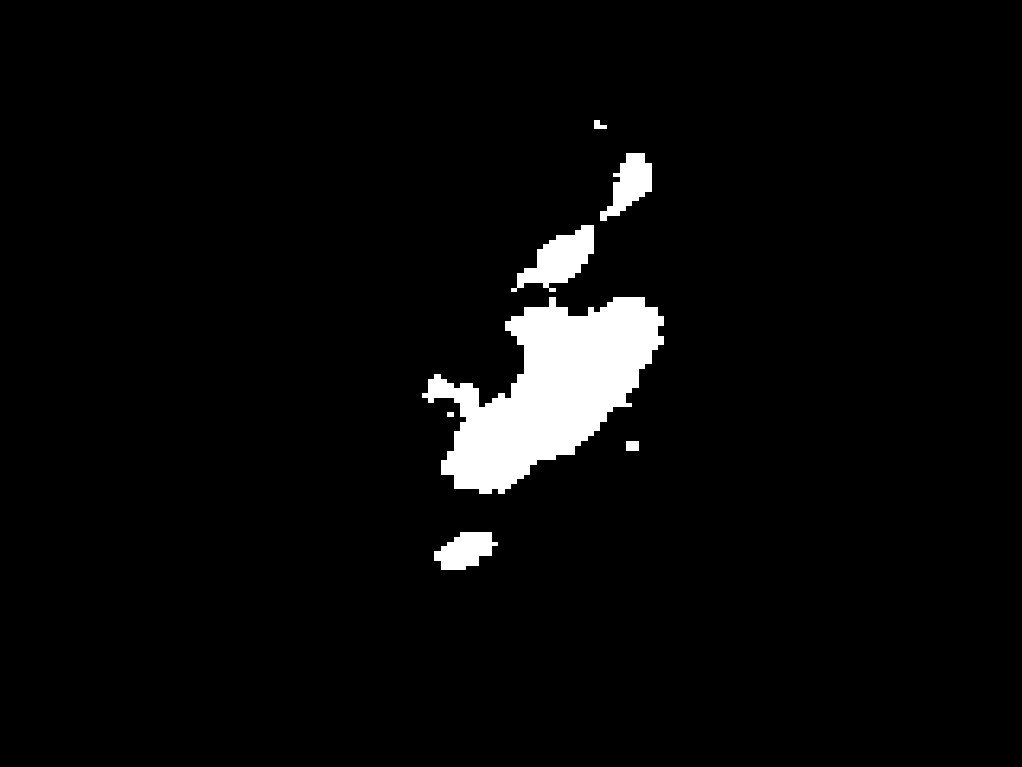}}
    \subcaption{globules}
    \end{subfigure}}\hfill
    \mbox{\begin{subfigure}[t]{0.2\textwidth}
    \centering\fbox{\includegraphics[width=\textwidth]{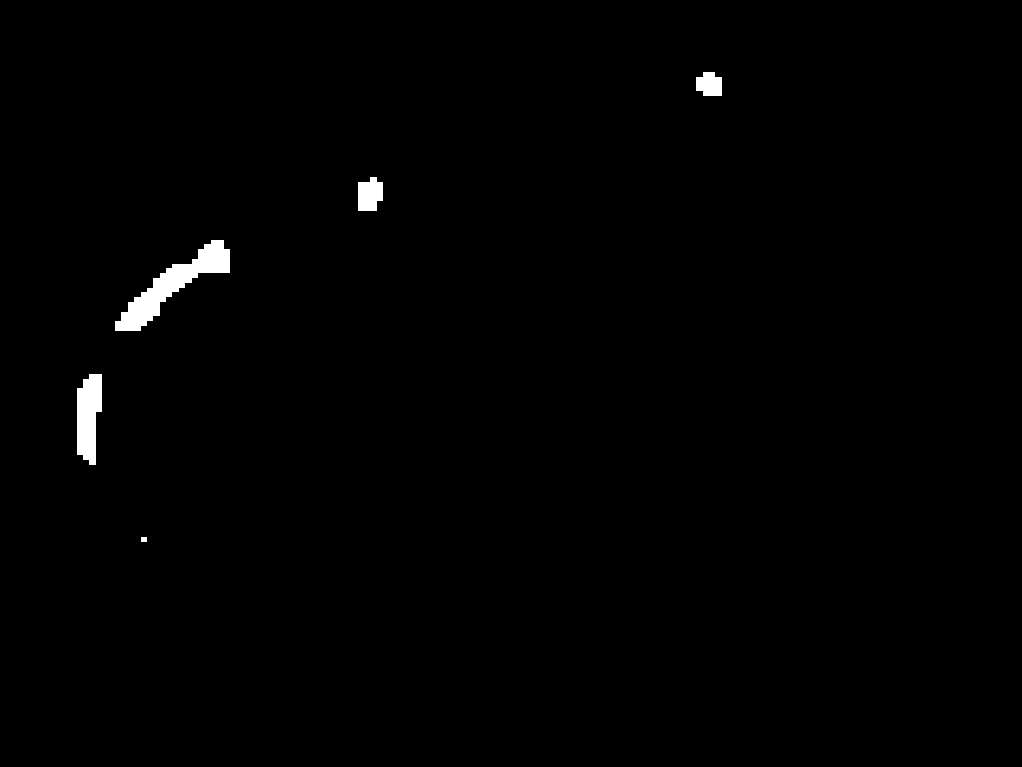}}
    \subcaption{streaks}
    \end{subfigure}}\hfill
    \hfill
    
    \hfill
    \mbox{\begin{subfigure}[t]{0.2\textwidth}
    \centering\fbox{\includegraphics[width=\textwidth]{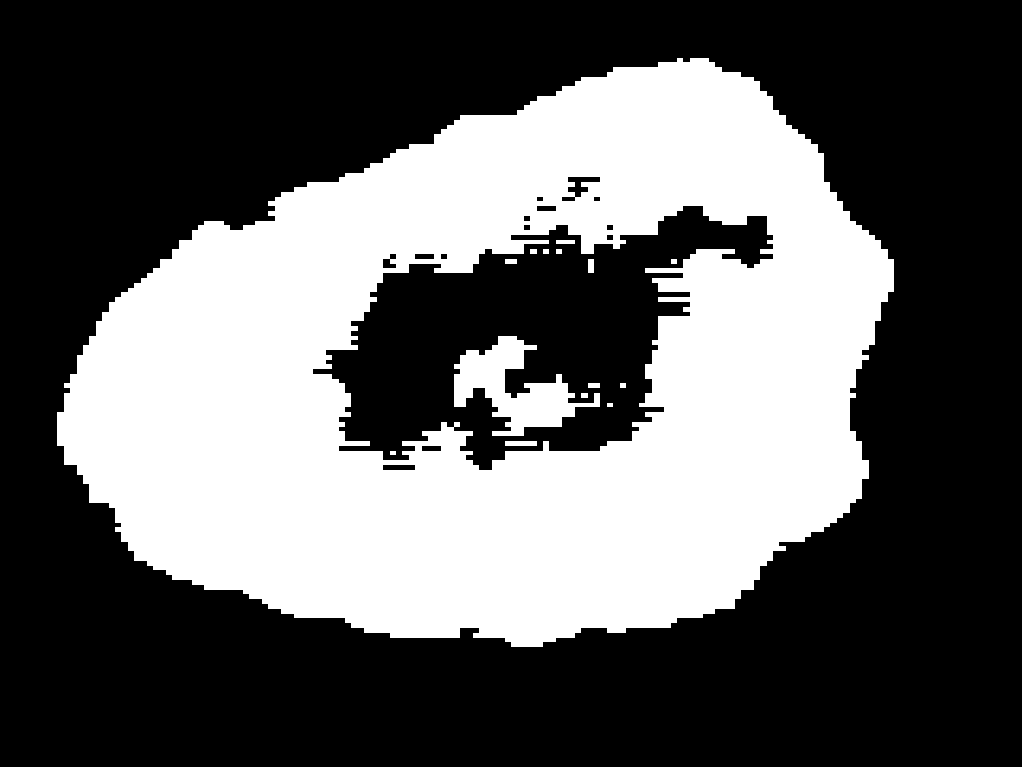}}
    \subcaption{pigment network}
    \end{subfigure}}\hfill
    \mbox{\begin{subfigure}[t]{0.2\textwidth}
    \centering\fbox{\includegraphics[width=\textwidth]{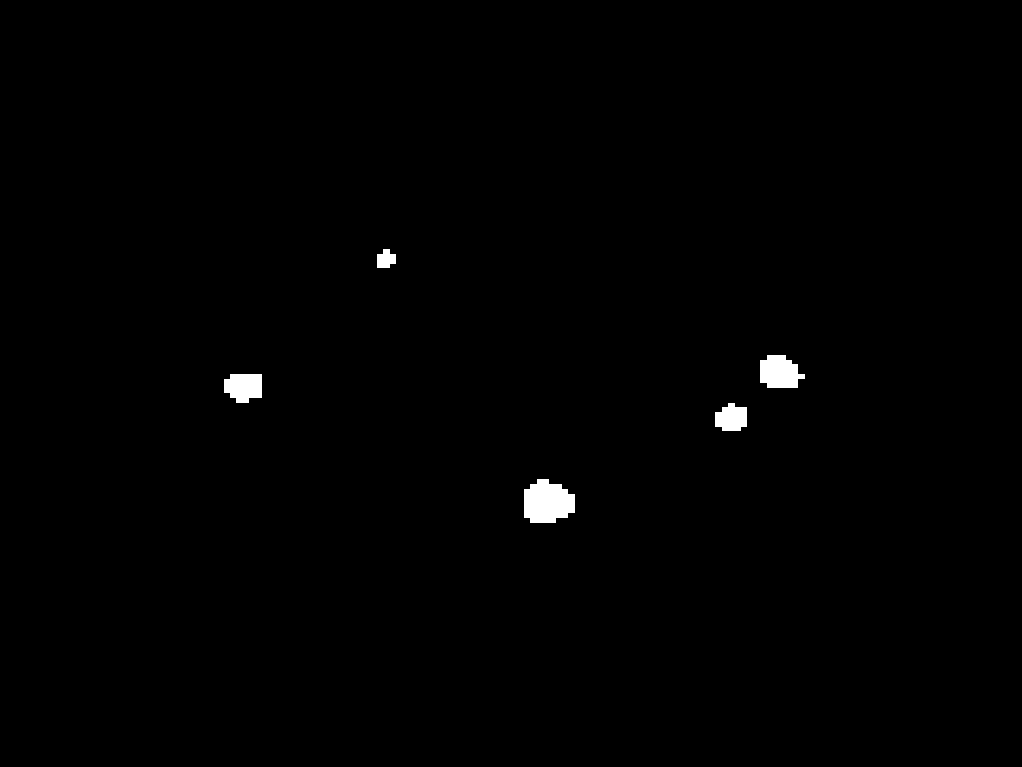}}
    \subcaption{milia-like-cyst}
    \end{subfigure}}\hfill
    \mbox{\begin{subfigure}[t]{0.2\textwidth}
    \centering\fbox{\includegraphics[width=\textwidth]{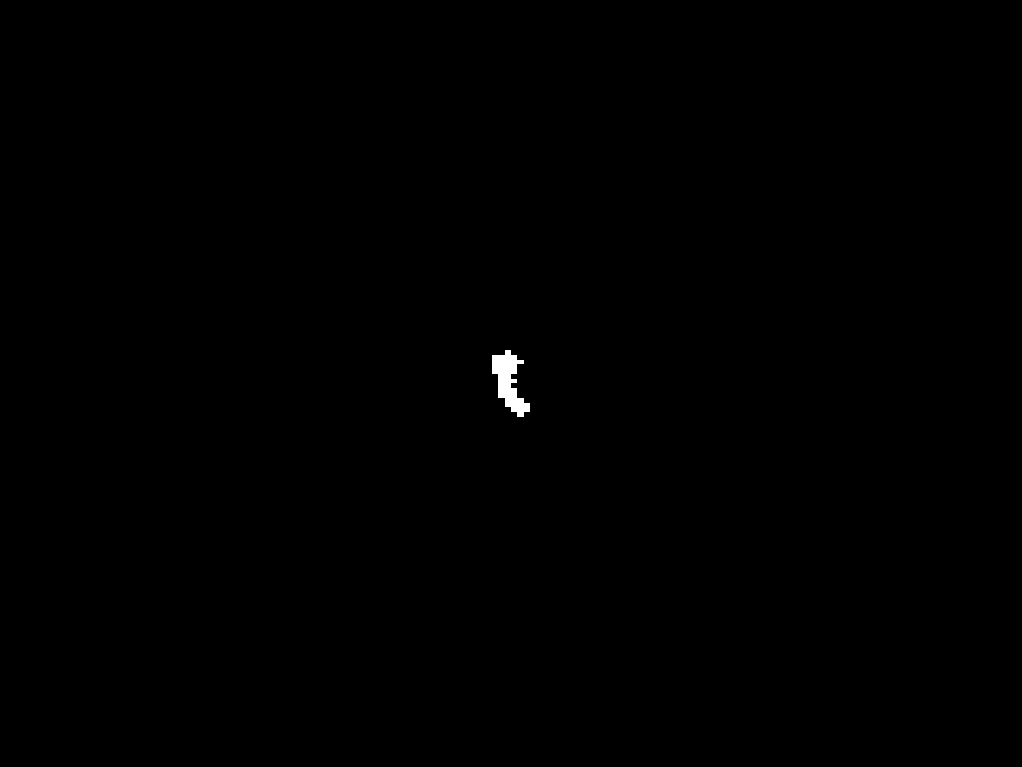}}
    \subcaption{negative network}
    \end{subfigure}}\hfill\hfill
    
    \caption{Features of an image} \label{fig:dl_features}
\end{figure}

Feature extraction is the task of identifying the areas of a skin lesion presenting the visual patterns of specific dermoscopic features.
In the Skincare project, the identification of each of the features has been implemented using a U-net \cite{navab_u-net:_2015} (de-) convolutional neural network on the dataset provided for the Task 2 of the ISIC 2018 challenge\footnote{\url{https://challenge2018.isic-archive.com/task2/}}.
The Skincare platform currently recognises five type of features: globules, streaks, pigment network, milia-like cyst, and negative network (see figure \ref{fig:dl_features}). Each of the five attributes is identified by a dedicated neural network. Hence, the same pixel might be identified by the platform as pertaining to more than one category.

\begin{figure}
    \hfill
    \mbox{\begin{subfigure}[t]{0.2\textwidth}
    \centering\fbox{\includegraphics[width=\textwidth]{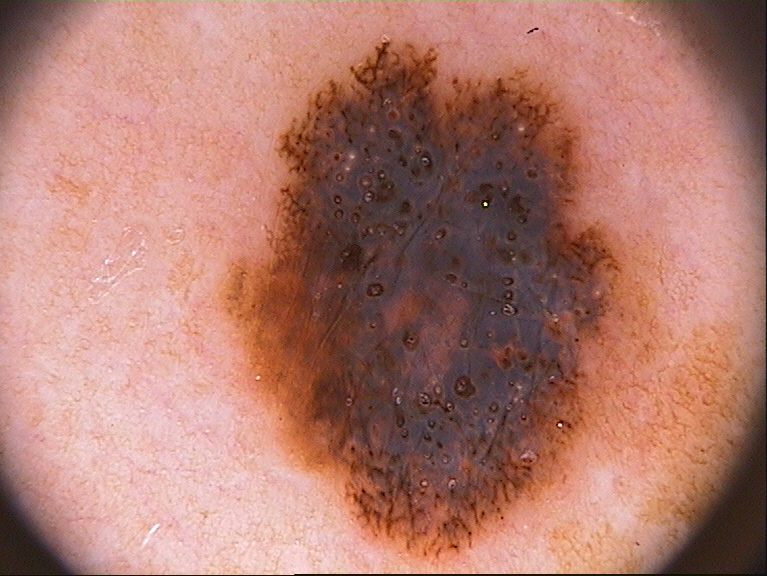}}
    \subcaption{original}
    \end{subfigure}}\hfill
    \mbox{\begin{subfigure}[t]{0.2\textwidth}
    \centering\fbox{\includegraphics[width=\textwidth]{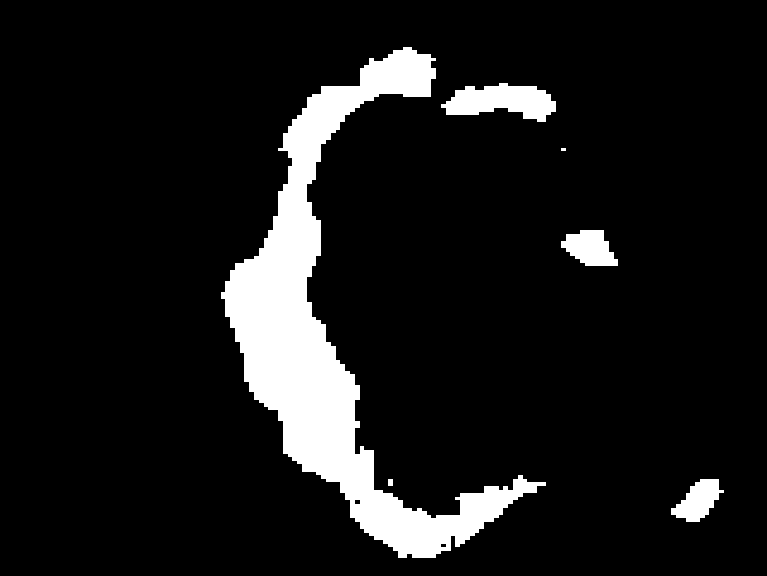}}
    \subcaption{pigment network mask}
    \end{subfigure}}\hfill
    \mbox{\begin{subfigure}[t]{0.2\textwidth}
    \centering\fbox{\includegraphics[width=\textwidth]{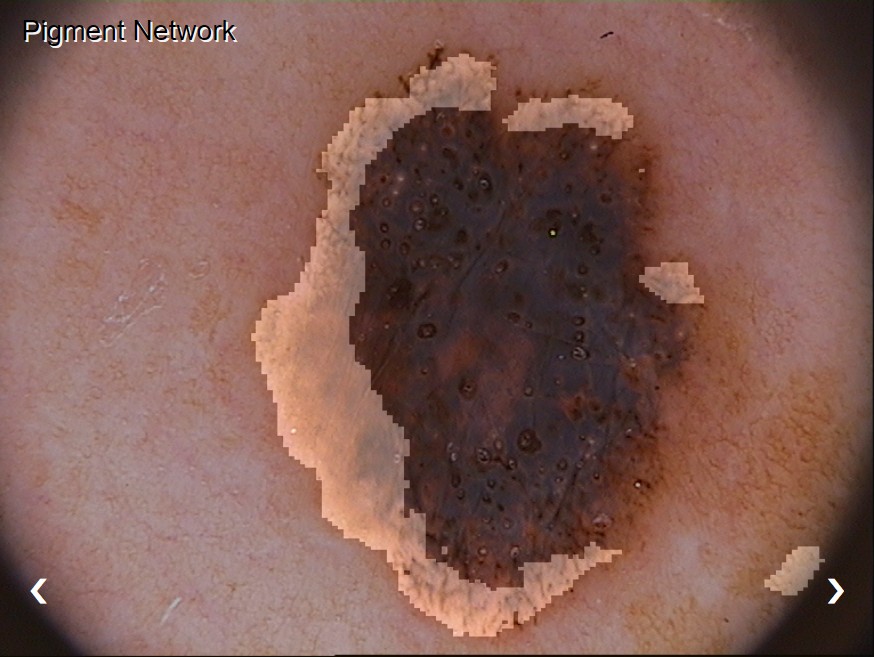}}
    \subcaption{pigment network  visualised as overlay (0.7)}
    \end{subfigure}}\hfill
    \mbox{\begin{subfigure}[t]{0.2\textwidth}
    \centering\fbox{\includegraphics[width=\textwidth]{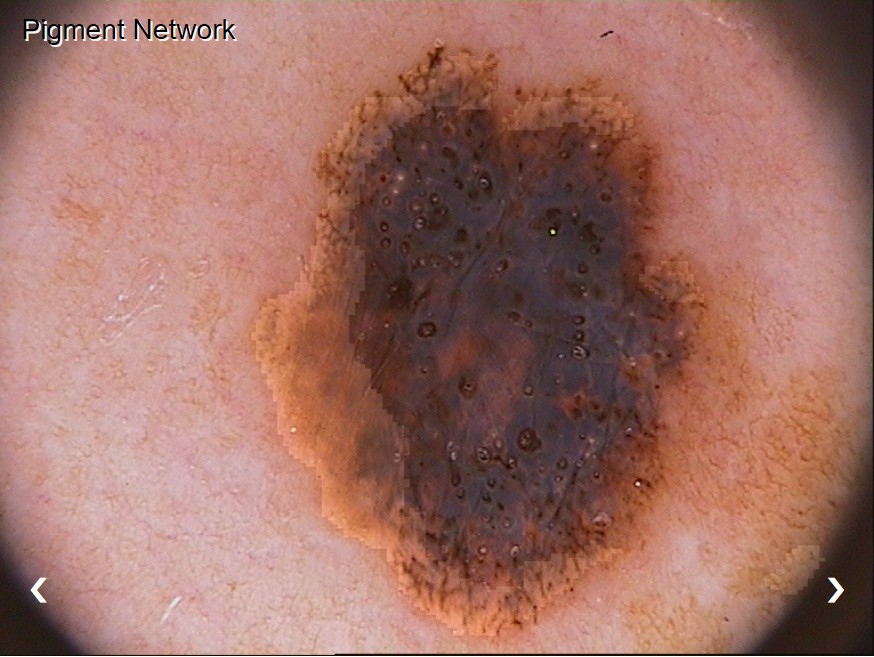}}
    \subcaption{pigment network  visualised as overlay (0.9)}
    \end{subfigure}}\hfill\hfill
    
    \hfill
    \mbox{\begin{subfigure}[t]{0.2\textwidth}
    \centering\fbox{\includegraphics[width=\textwidth]{images/original2.jpg}}
    \subcaption{original}
    \end{subfigure}}\hfill
    \mbox{\begin{subfigure}[t]{0.2\textwidth}
    \centering\fbox{\includegraphics[width=\textwidth]{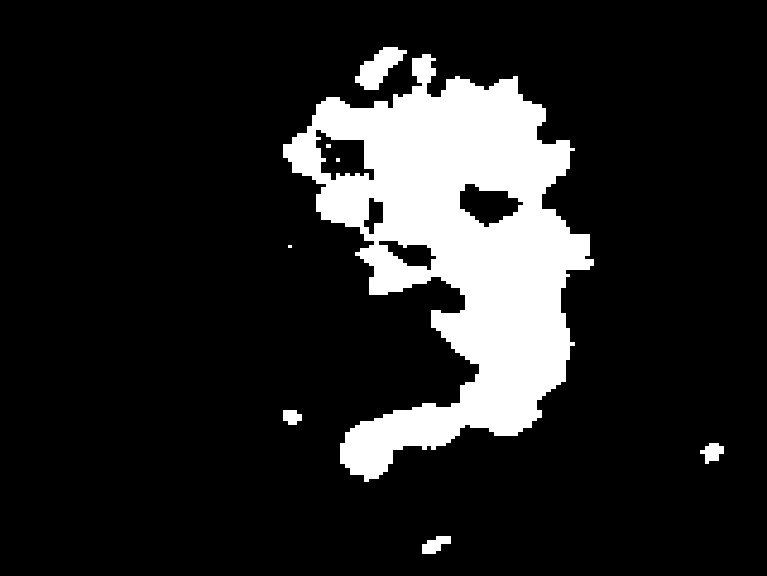}}
    \subcaption{globules mask}
    \end{subfigure}}\hfill
    \mbox{\begin{subfigure}[t]{0.2\textwidth}
    \centering\fbox{\includegraphics[width=\textwidth]{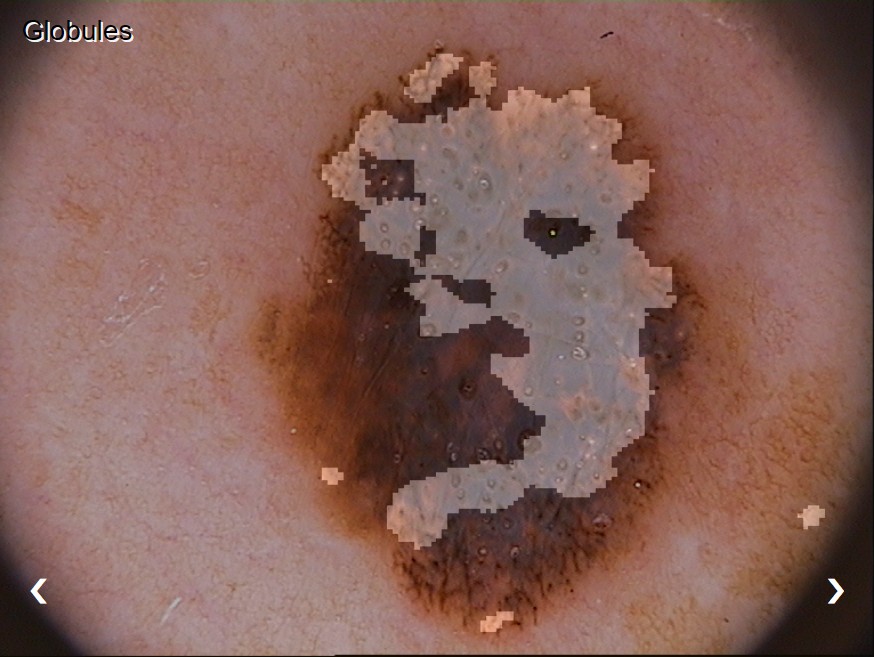}}
    \subcaption{globules visualised as overlay (0.7)}
    \end{subfigure}}\hfill
    \mbox{\begin{subfigure}[t]{0.2\textwidth}
    \centering\fbox{\includegraphics[width=\textwidth]{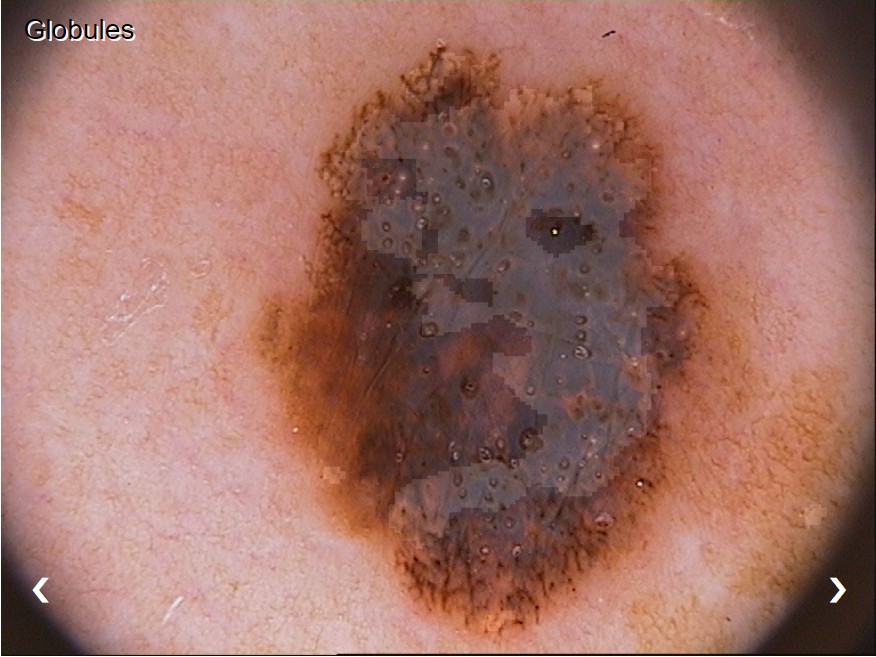}}
    \subcaption{globules visualised as overlay (0.9)}
    \end{subfigure}}\hfill
    \hfill
    
    \caption{Advantage of overlays: transparency can be controlled by user, here: 0.7 and 0.9} \label{fig:overlays}
\end{figure}

The current implementation computes area masks of resolution 180x180 pixels, which are then re-scaled to the resolution of the original image for overlapping and interactive masking. The advantage of overlays, whose transparency can be easily configured via a slider, is shown in figure \ref{fig:overlays}.

\newcommand{\tm}{\textsuperscript{TM}}

\subsection{Feature Extraction (visual filter-oriented colour analysis)}
The \textit{Asymmetry} feature stems from the fact that lesion images  are generally not symmetric with the major x–y axes of the images. However, to judge if there is any asymmetry in shape, the lesion axes must be aligned to the major axes of the image. To first accomplish this alignment, it is necessary to transform the segmented image  by finding the lesion’s minimum enclosing rectangle and extracting the rectangular matrix from the image. This matrix provides the major and minor axes, along with the tilt angle of the rectangle. Next, we calculate the rotation matrix from the tilt angle and transform the segmented image to the rotated image.
The asymmetry score is calculated from a total of eight parameters. The first two parameters, vertical and horizontal asymmetry, are calculated by overlapping the binary form of the warped segmented image with the mirror images in vertical and horizontal directions. The sum of all the non-zero pixels in the image is computed and the asymmetry level (AS) is calculated as a percentage of these non-zero pixels in the overlapped image over the lesion area.
The remaining six parameters refer to the asymmetry in structure and are calculated as the distance between the lesion centroid and the weighted centroids of the color contours (obtained from the colour variegation feature).
The \textit{Border irregularity} feature is generally defined as the level of deviation from a perfect circle and measured by the irregularity index. The minimum value of the irregularity index  corresponds to a perfect circle. 
The \textit{Diameter greater than 6 mm} feature refers to the size of a lesion. The diameter of the lesion is calculated from the side length of the minimum area rectangle and the conversion factor from pixels to millimeters.
The \textit{Color variegation} feature indicates the  number of  colours of the lesion from the HSV (Hue, Saturation, and Value) image by grouping all pixels with  HSV values within a given range.
The predefined color set includes the colours white, red, light brown, dark brown, blue-gray, and black. In general, a benign mole has one or two colours while a malignant mole may have more than three colours.

\begin{figure}[!htb]
   \hfill
   \mbox{\begin{subfigure}[t]{0.3\textwidth}
    \centering\fbox{\includegraphics[width=\textwidth]{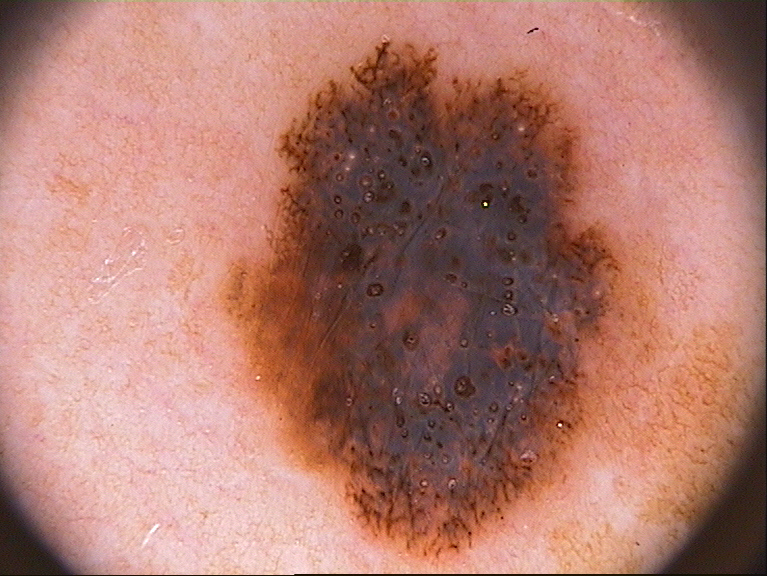}}
    \subcaption{original}\end{subfigure}}\hfill
    \mbox{\begin{subfigure}[t]{0.3\textwidth}
    \centering\fbox{\includegraphics[width=\textwidth]{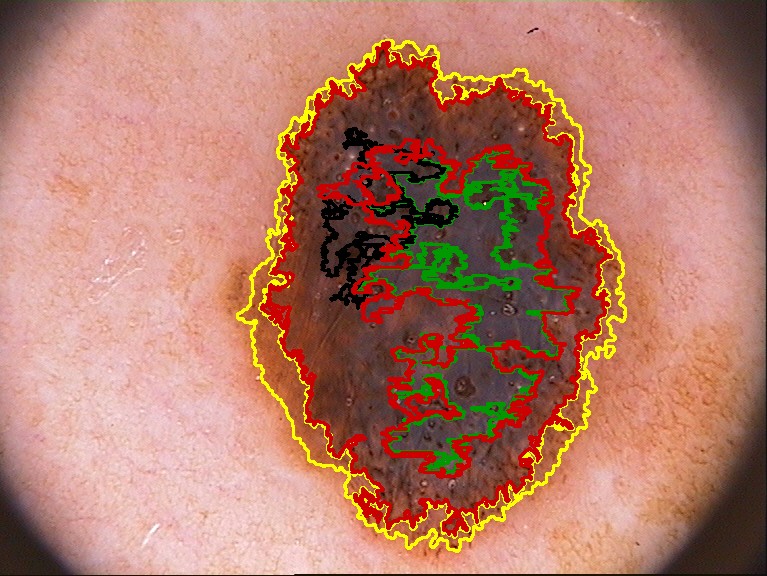}}
    \subcaption{colour regions}
    \end{subfigure}}\hfill\hfill
    \caption{Colour variegation: a) original image, b) image showing 4 different colour regions (see also figure \ref{fig:classification})}
    \label{fig:steps_colors}
\end{figure}

These (numerical) feature values are projected into a range of $[0..10]$ and  displayed in the web page (see the blue bars in figures \ref{fig:classification} and \ref{fig:classificationtouch}) as
\textbf{A1/A2}: horizontal resp. vertical asymmetry, 
\textbf{B}: irregularity index of the border, 
\textbf{D1/D2}: the diameter values (horizontal resp. vertical).
The colour variegation is directly visualised as one of four images shown in the web page (see figure \ref{fig:classification} for more details of the presentation in the web page).

\subsection{The ABCD Rule}

Melanoma detection can be done following the so-called ABCD rule which stands for Asymmetry, Border irregularity, Color variegation, and Diameter greater than 6 mm.
The ABCD rule is well known and described in various scientific articles, e.g. \cite{friedman1985ABCD, she2007combination, jain2015computer}.
According to \cite{harrington2017diagnosing}, the ABCD rule is the most popular and most effective algorithm for ruling out melanoma among all the computerised methods for melanoma detection. There are other prominent clinical methods for early melanoma detection. The ABCDE method extends the ABCD rule by adding the evolution of the lesion (e.g., elevation, enlargement, and color change). Although the ABCDE rule has been validated in clinical practice, no randomised clinical trials have shown that there is an improvement in the early detection of melanoma \cite{45pmid25698455, 48pmid16618863}.

Colour, architectural disorder, symmetry, and homogeneity/heterogeneity of mole structures are used in the 
in  the  C.A.S.H. algorithm \cite{42pmid17190620} (which has a lower specificity compared to the ABCD rule \cite{8pmid20671054, 43pmid26601859}.) Video microscopy and epifluorescence are used for pattern analysis to categorise the type of skin lesion based upon its general appearance, border,  surface, pigmented patterns, and depigmentation \cite{39SaezAchaSerrano, 40pmid3668002}. 
The Glasgow seven-point checklist performs diagnosis on three major features (change in size of lesion, irregular pigmentation, and irregular border) and four minor features (inflammation, itching sensation, diameter greater than 7 mm, and oozing of lesions) \cite{1pmid23063256, Scharcanski2013CVT, 8pmid20671054, 45pmid25698455, 46pmid15585738}. Because of its inherent complexity, it is not widely adopted and has a lower pooled sensitivity than the ABCD rule \cite{47pmid28264830}. Image acquisition methods have been developed to differentiate the amount of light absorbed, transmitted, or back-scattered by the melanin part of a lesion. Examples of such methods are reflectance confocal microscopy, hyperspectral imaging, and optical coherence tomography \cite{43pmid26601859}. However, these methods aren't standardised yet to accurately calibrate  absorbance and reflectance from the image. Ultrasound technology and electrical bioimpedance measurements are used to retrieve information about the inflammatory process in the skin. However, these ultrasound images are difficult to interpret and the electrical impedance of the skin can vary greatly based on age, gender, and body location \cite{8pmid20671054}.

Pattern analysis and the ABCD rule are the oldest and most widely adopted methods for melanoma detection \cite{41pmid28109068} by dermatologists; therefore is is very suitable for building an automatic AI-based system for it. The SkinCare system includes and extends the software described in  \cite{DBLP:journals/symmetry/KalwaLKP19} which is available on github\footnote{\url{https://github.com/ukalwa/melanoma-detection-python}}. The software uses the SVM classification algorithm and was trained on the PH2 data set \cite{mendoncca2013ph}.

In order to further support the medical judgement of an image, the user interface includes a segmentation based on CNN or filter-based computer vision, relevant colour areas (so called centroids), the graphics that visualise the asymmetry values and numeric values for asymmetry, borders, and diameters (second row of images titled `ABCD Features' and the information left of the large image in figure \ref{fig:classification}). More information about the user interface is provided in section \ref{sec:user-interfaces}.

\section{Architecture}
\label{sec:architecture}

\begin{figure}
\centering
\includegraphics[width=\linewidth]{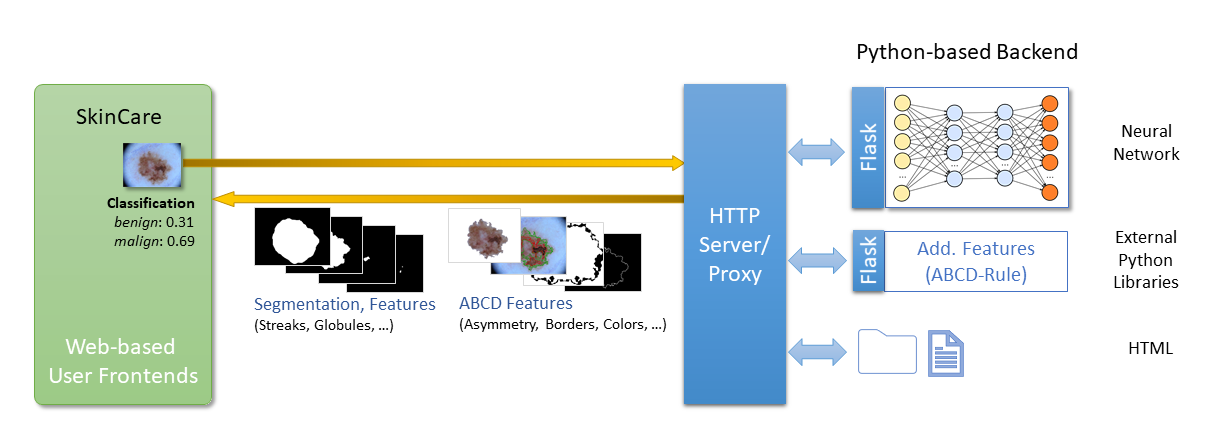}
\caption{Skincare Basic Architecture: The web page interacts with the HTTP server, that integrates proxy entries for the different flask servers in the backend.}
\label{fig:architecture}
\end{figure}

Figure \ref{fig:architecture} shows the basic architecture of the Skincare system.

The back-end consists of an HTTP server that delivers the static HTML pages and interacts with the flask\footnote{\url{http://flask.palletsprojects.com/}} servers wrapping the classification networks and other Python libraries. This ensures maximum flexibility in adding/exchanging computational services. 
The front-end consists of several web pages for the intelligent user interfaces which are described in detail in the next section.
In the future, more multimodal interfaces \cite{mmbook1} can be built with the help of domain-specific speech-based dialogue systems \cite{sonntag2009supporting}.

\section{Intelligent User Interfaces}	
\label{sec:user-interfaces}

\begin{figure}
\centering\captionsetup{width=.8\linewidth}
\fbox{\includegraphics[width=.8\linewidth]{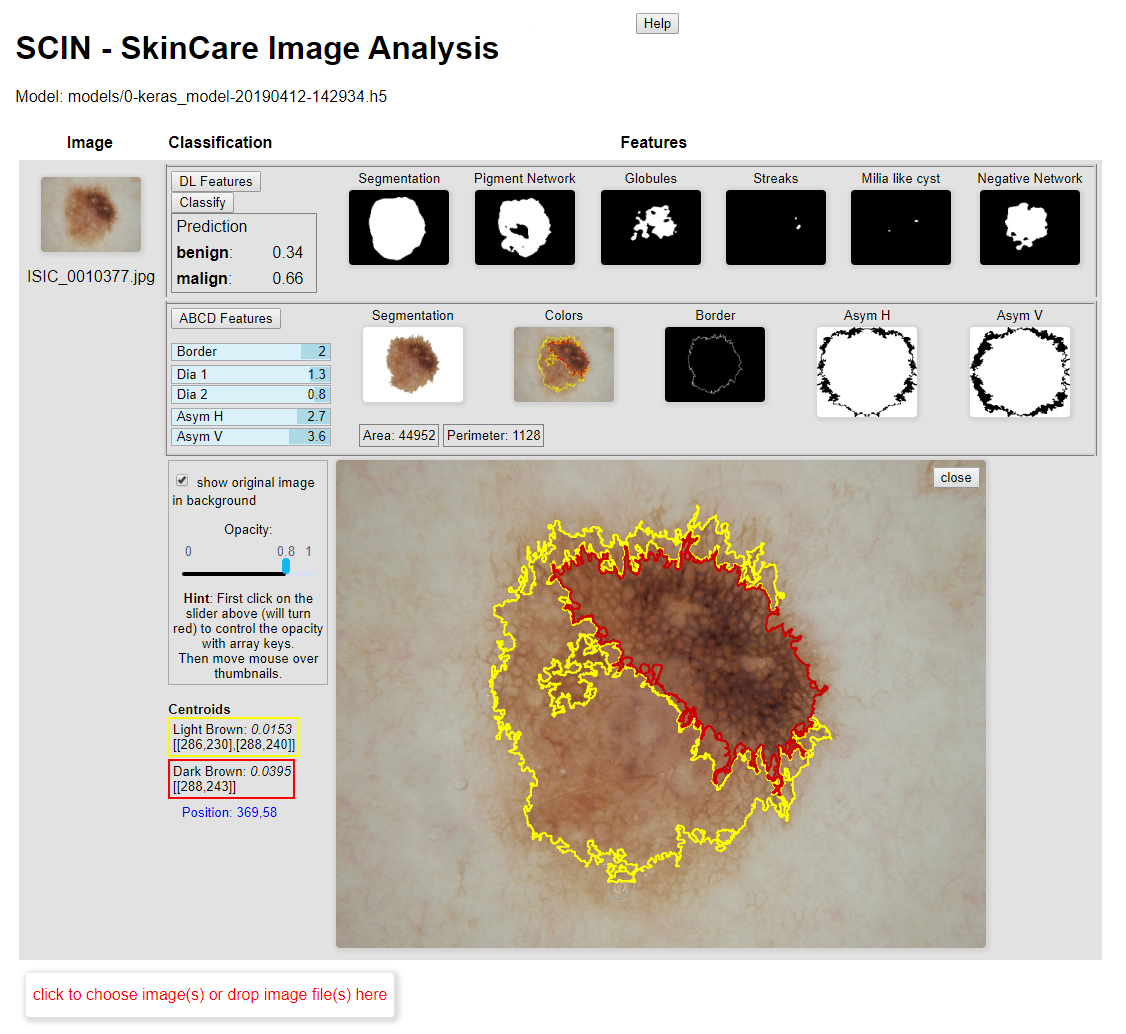}}
\caption{The classification web page (decision support). The first row shows classification results (benign/malign) and extraction features computed by the deep learning algorithm (`DL Features') such as globules and streaks. The second row (`ABCD Features') displays features computed by computer vision methods and some numeric values categorising attributes like border, diameter, and asymmetry (blue bars). The large image is selected by mouse-over (here showing colours from the ABCD features) and is rendered as an overlay to the original image. The shown large image has three color regions, namely a region of dark-brown, marked red, and two regions of light-brown, marked yellow. More information about the centroids of these regions is displayed in the yellow and red boxes, respectively.}
\label{fig:classification}
\end{figure}

\begin{figure}
\centering\captionsetup{width=0.95\linewidth}
\includegraphics[width=1.0\linewidth]{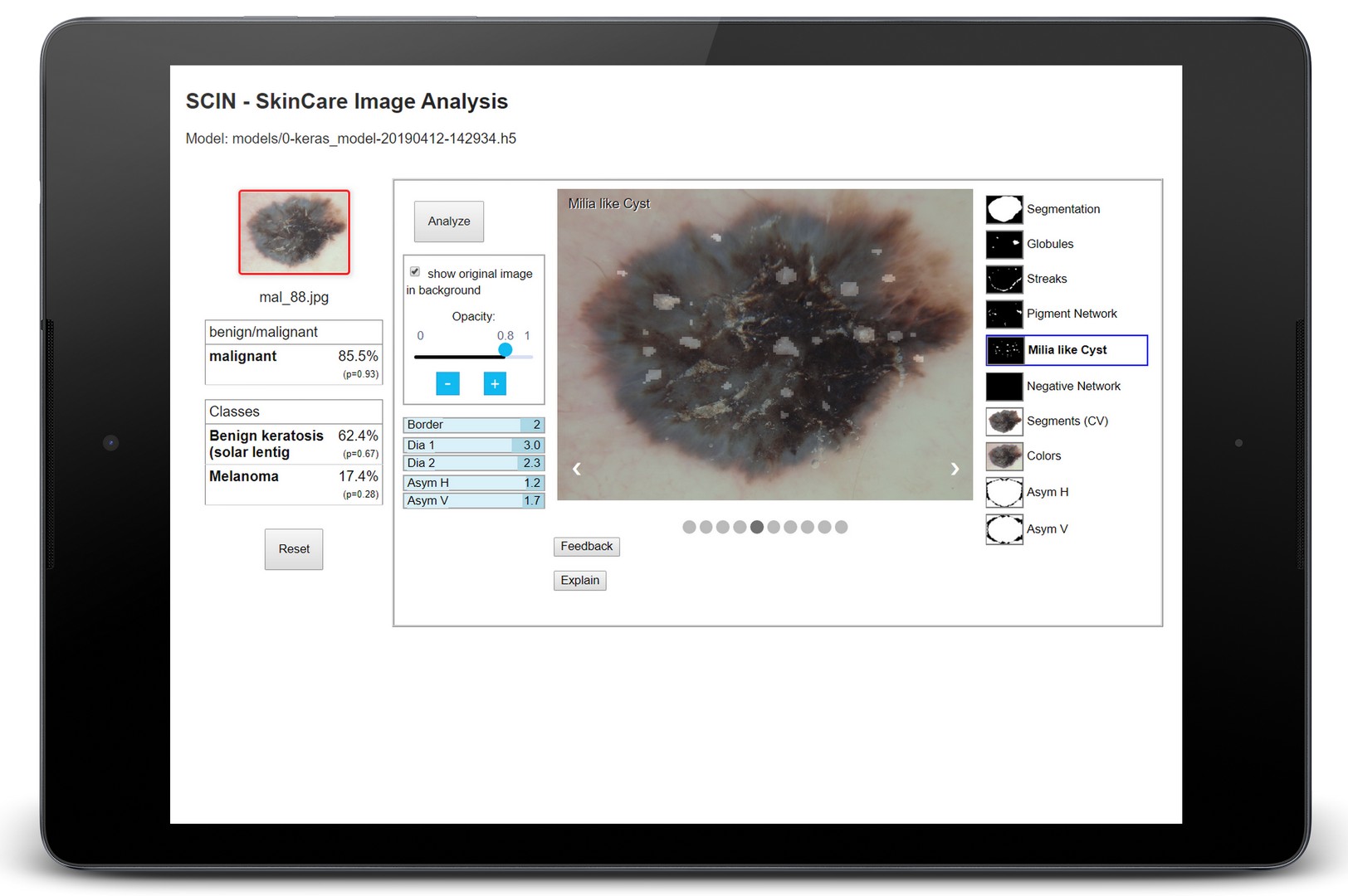}
\caption{The touch-optimised version of the classification web interface.  The original image in the middle has an overlay with the selected `Milia like Cyst' mask (opacity 0.8, which can be changed via the blue slider controls).  The left part of the page shows the original image with a color-coded border (representing the probability of malignancy) and the numerical results (confidences) of the binary and the 8-class classifier (the probability is shown in brackets).}
\label{fig:classificationtouch}
\end{figure}

\begin{figure}
\centering\captionsetup{width=0.9\linewidth}
\fbox{\includegraphics[width=0.9\linewidth]{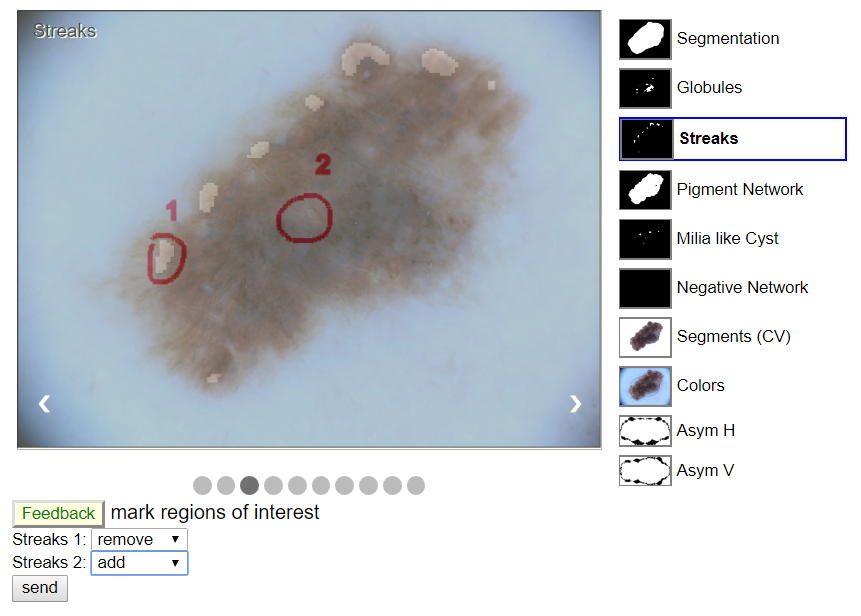}}
\caption{User feedback, Interactive Machine Learning (IML): The user marked two regions of interest on the `Streaks' mask and chose actions (add/remove). Clicking on the `send' button will post the data to the back-end where it can be used as feedback to the machine learning process. See \url{http://iml.dfki.de} for further examples of how IML is performed.}
\label{fig:feedback}
\end{figure}

\begin{figure}
	\centering\captionsetup{width=0.7\linewidth}
	\fbox{\includegraphics[width=0.7\linewidth]{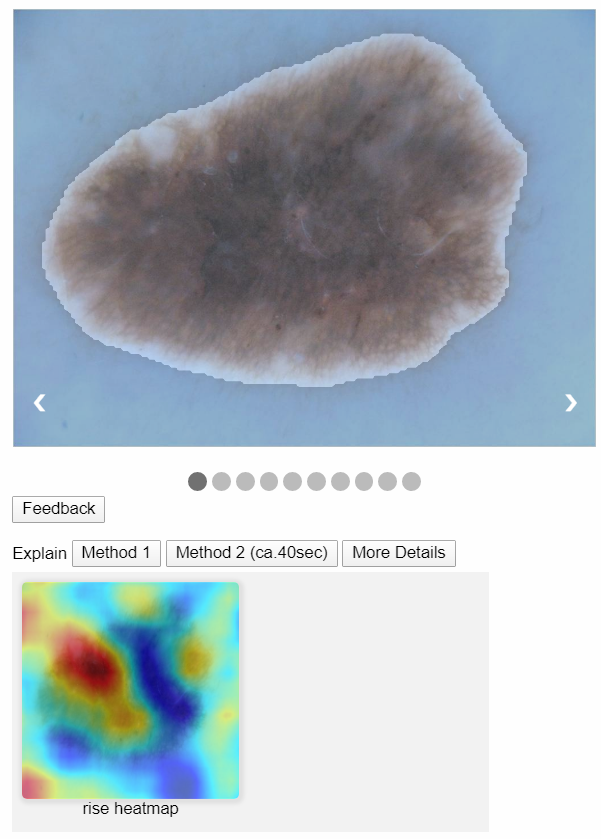}}
	\caption{RISE: The image on the bottom shows the result of the RISE algorithm called with 100 masks (triggered over the button ``Explain - Method 2''). ``Method 1'' uses the GradCAM algorithm, ``More Details'' opens another pane where the parameters can be changed individually, as shown in figure \ref{fig:explanation_gradcam}. For more details on this method see section \ref{sec:objectives}.}
	\label{fig:explanation_rise}
\end{figure}

\begin{figure}
	\centering\captionsetup{width=0.7\linewidth}
	\fbox{\includegraphics[width=0.7\linewidth]{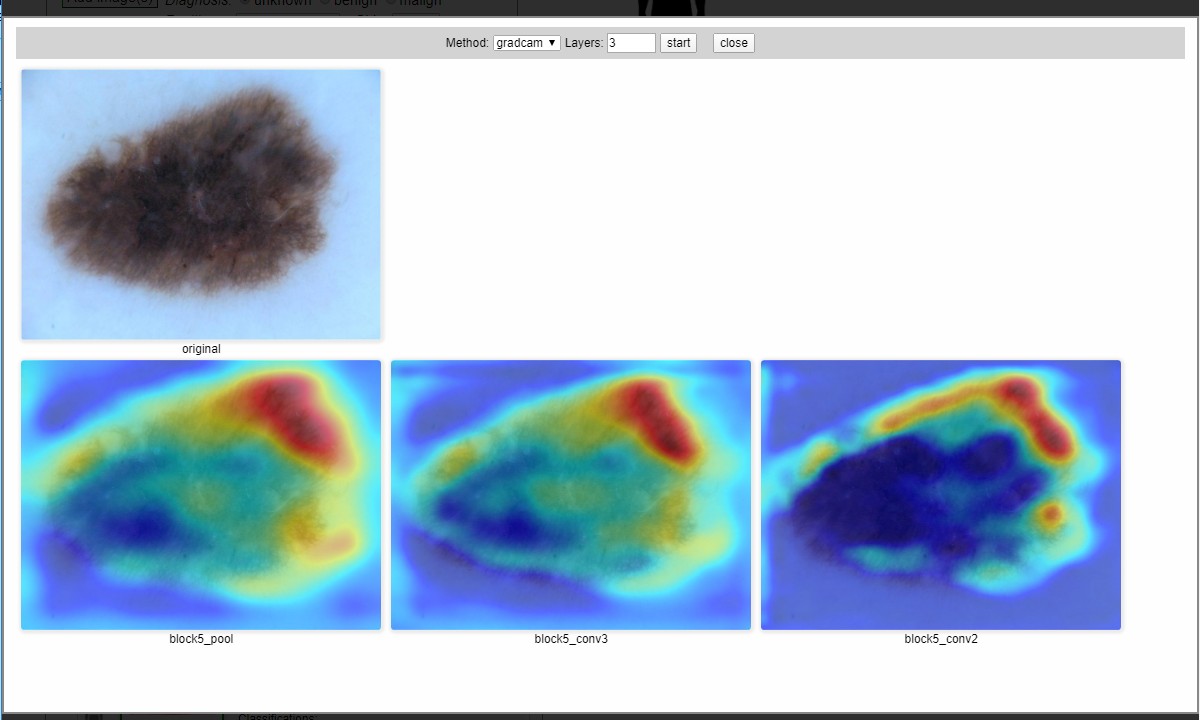}}
	\caption{GradCAM: The three images at the bottom show heatmaps of the last layers of the model. This is a more elaborated pane for experienced users where the methods and parameters can be changed directly. For more details on this method see section \ref{sec:objectives}.}
	\label{fig:explanation_gradcam}
\end{figure}

\begin{figure}
\centering\captionsetup{width=.8\linewidth}
\fbox{\includegraphics[width=.8\linewidth]{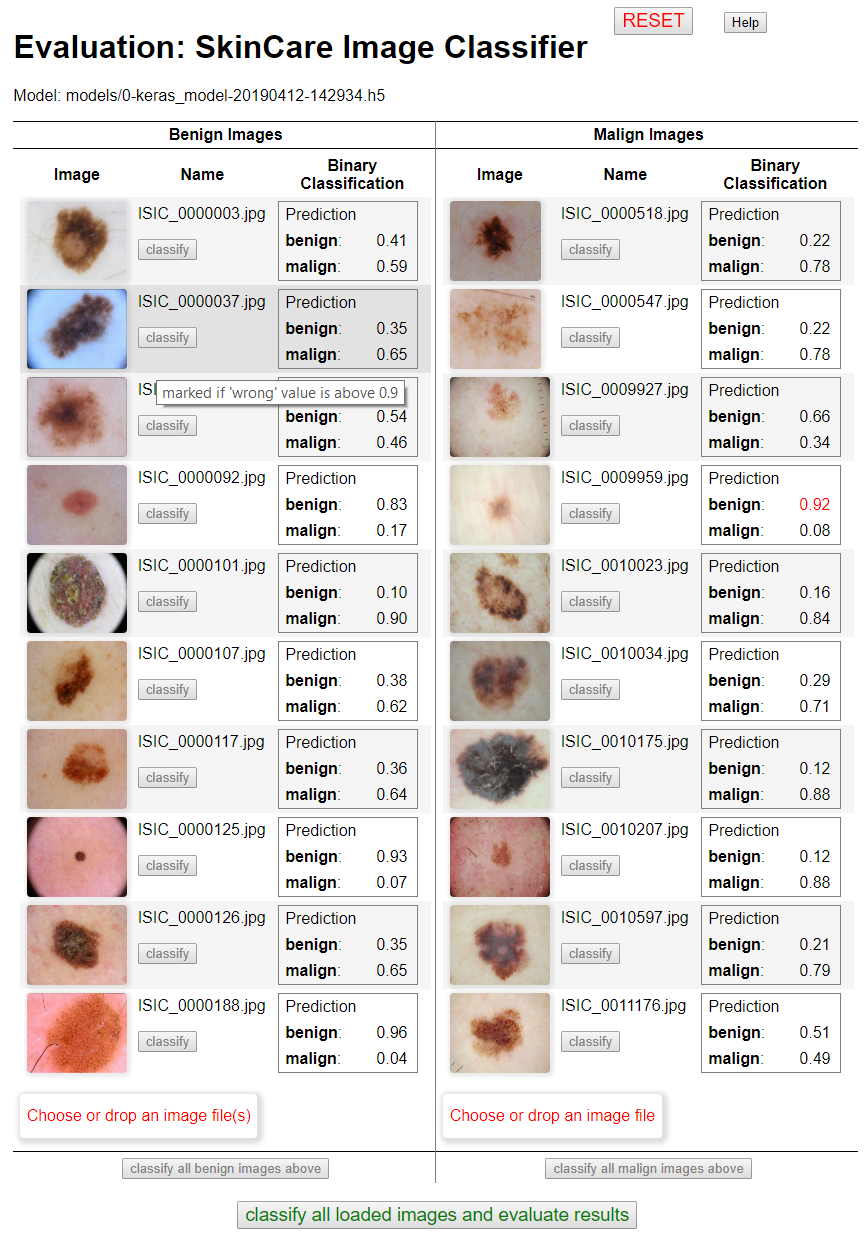}}
\caption{The evaluation base: two sets of dermatological images (benign, left column, and malign, right column) can be uploaded; the classification results are shown in this part of the page. The page also contains some quality measures (shown in figure \ref{fig:evaluationresults}).}
\label{fig:evaluation}
\end{figure}

The Skincare graphical user interfaces comprises of three web pages:
\begin{description}
    \item[Classification page (decision support):] (\url{http://www.dfki.de/skincare/classify.html}) A page to upload images and start different image analysis processes (`Deep Learning features', classification and `ABCD Features' (see figure \ref{fig:classification}). The different masks (for segmentation, globules, colours, for example) can be used as overlays over the original image to allow for further inspection by the user (selected by mouse-over). Results of the ABCD module are integrated as additional masks (for segmentation, colours, and asymmetry) and (blue) bars representing the numerical values. 
        
    \item[Extended classification page for touch screens:] (\url{http://www.dfki.de/skincare/classifytouch.html}) A variant of the classification page that is optimized for touch screens (see figure \ref{fig:classificationtouch}). This presentation is more compact to support smaller screens and the selection of masks can also be done by swiping through the graphics. Some additional functionalities like two kinds of classification (binary and 8-class), color coding showing the probability of malignancy, user feedback (figure \ref{fig:feedback}) and heatmaps (figures \ref{fig:explanation_rise} and \ref{fig:explanation_gradcam}) of the last layers of the model are included in this version. 
    
    \item[Evaluation page:] (\url{http://www.dfki.de/skincare/evaluate.html}) This page allows us to evaluate the (binary) classifier results. Two sets of images with known results (benign and malign) can be uploaded and classified. All images are classified and the evaluation measures precision, recall, specificity, accuracy, F1, FPR (false-positive-rate), and TPR (true-positive-rate) are presented\footnote{generated with \url{www.highcharts.com}} with different thresholds. Additional charts show the ROC (receiver operating characteristic) curve and the relation between precision and recall (figure \ref{fig:evaluation}). 
\end{description}

\begin{figure}
\centering\captionsetup{width=.7\linewidth}
\fbox{\includegraphics[width=.7\linewidth]{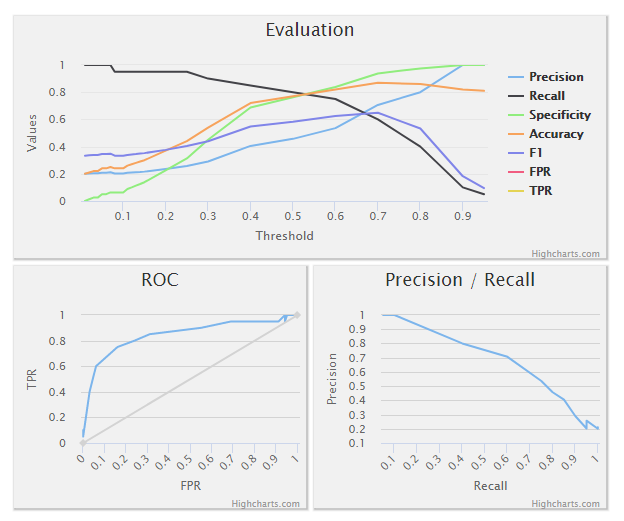}}
\caption{The evaluation results: the system computes several evalutaion measures based on the classification results (see figure \ref{fig:evaluation}) and displays them in corresponding graphical charts.}
\label{fig:evaluationresults}
\end{figure}

\section{REST API}
\label{API}

The Skincare implementation provides a web REST API to perform classification, segmentation, and feature extraction. The calls return either a JSON dictionary or an image. In the following, the REST commands available from the classification server are described in detail.

\subsection{Info}
Returns information about the used classifier.

\begin{description}[leftmargin=\parindent,labelindent=\parindent]
    \item[path] `/model\_info'
    \item[method] POST
    \item[input parameters] none
    \item[returns] `binary\_classification\_model': The model used for this classification
\end{description}

\subsection{Static content}
Used to serve static html pages, useful to enrich the service.
Just create a subdirectory named `html' in your server working directory and store your files there.
\begin{description}
    \item[path] `/html/<file>'
    \item[method] POST
    \item[input parameters] the name of the file to retrieve
    \item[returns] the specified file, contained in the `html' subdirectory
    \item[Content-Type] inferred from the file extension
\end{description}

\subsection{Classification: Benign vs. Malignant}

Binary classification is the process of analyzing an image of a lesion and return the probability distribution between \textit{benign} (0) and \textit{malign} (1).

\begin{description}
    \item[path] `/classify/binary'
    \item[method]  POST
    \item[input parameters]: `file`: the image to classify. Must be a JPEG or PNG image. No alpha.
    \item[returns] a JSON structure with info about the classification with the following fields:
    \begin{description}
     \item[error] If there was an error, otherwise this entry is absent
     \item[filename] The name of the file provided as input
     \item[prediction] A 2-dimension array with the probability for benign and malignant cases, respectively.
    \end{description}
\end{description}

\subsection{Segmentation}

Segmentation is the process of taking the image of a lesion as input and generating another image representing a binary \textit{mask} containing the lesion.

The output image is a gray-scale PNG image, pixels can be only black (the pixels are outside of the lesion) or white (the pixels pertains to the lesion).

\begin{description}
    \item[path]  /segment
    \item[method]  POST
    \item[input parameters]: `file': the image to segment. Must be a JPEG or a PNG image. No alpha.
    \item[returns] depending on result:
    \begin{itemize}
        \item In case or error, returns a JSON file with the error reason, \\
        Content-Type: `application/JSON'\\
        `error`: The reason of the error
        \item Otherwise, returns a PNG image (same size as input), gray-scale (1 channel), with the lesion mask.\\
        Content-Type: `image/png'
    \end{itemize}
\end{description}

\subsection{Feature Extraction}

Feature Extraction is the process of extracting, from the image of a lesion, pixel areas classified as pertaining to a certain category.

The output image is a gray-scale PNG image, pixels can be only black (the pixels are not in the feature class) or white (the pixels pertains to the feature class).

\begin{description}
    \item[path]  /extract\_feature/<feature\_class>\\
        with `feature\_class' one of: `globules', `streaks', `pigment\_network', `milia\_like\_cyst', `negative\_network'.
    \item[method] POST
    \item[input parameters]: `file': the image to process. Must be a JPEG or a PNG image. No alpha.
    \item[returns] depending on result:
    \begin{itemize}
        \item In case or error, returns a JSON file with the error reason, \\
        Content-Type: `application/JSON'\\
        `error`: The reason of the error
        \item Otherwise, returns a PNG image (same size as input), gray-scale (1 channel), with the lesion mask.\\
        Content-Type: `image/png'
    \end{itemize}
\end{description}

\section{Installation}

Download the main package for the Skincare classification project.
It contains a number of files, put all of them in an \textit{empty} directory:

\begin{description}
    \item[`README.md'] containing the installation instructions
    \item[`skincare\_dfki-x.y.z-py3-none-any.whl'] The installable python package.
    \item[`models'] directory with the binary trained models.\\
    E.g.  `0-keras\_model-20190412-142934.h5` is the trained model for binary classification.\\
     `model-segment\_weights.h5` is the model for the segmentation.
     \item[`skincare\_config.json`] The configuration file for the server. Edit it to point to new models, if needed.
     \item[`REST-API.md`] Documentation for the REST API (see also previous section)
     \item[plus] some example images for testing.
\end{description}

\subsection{Install Python Package}

Create a Python3 environment and install the package from the wheel archive
\begin{verbatim}
bash
cd path/to/your_directory
python3 -m venv skincare-p3env
source skincare-p3env/bin/activate

pip install -U skincare_dfki-x.y.z-py3-none-any.whl    
\end{verbatim}

\subsection{Run the server providing the REST-API}

The http REST interface is implemented using Flask.
To run the server from a terminal:
\begin{verbatim}
bash
cd path/to/your_directory
export FLASK_APP=skincare.networking.__main__.py
python -m flask run
\end{verbatim}

By default, the server takes connections on port 5000.
From a browser, you can use the server with:

\begin{itemize}
    \item `http://127.0.0.1:5000/model\_info` returns info about the loaded classifiers.
    \item `http://127.0.0.1:5000/html/hello.html` Test if serving the static html pages works.
    \item `http://127.0.0.1:5000/classify/binary` performs the actual classification. This must be a POST, providing the image.

\end{itemize}

\noindent In order to support public connections and different ports:
\begin{verbatim}
python -m flask run --host=0.0.0.0 --port=80
\end{verbatim}

\section*{Acknowledgements}

We would like to thank all student assistants that contributed to the development of the platform: 
Abraham Ezema, Haris Iqbal, Md Abdul Kadir, and Duy Nguyen. This research is partly supported by H2020 and BMBF. See our project page about Skincare  \url{http://medicalcps.dfki.de/?page_id=1056} and the Interactive Machine Learning Lab (\url{http://iml.dfki.de}).

\bibliography{skincare,SkinCare2}

\end{document}